\documentclass[iop]{emulateapj}

\pdfoutput=1
\usepackage{natbib}
\usepackage{epsfig}

\newcommand{\figref}[1]{Figure \ref{#1}}
\newcommand{\eqref}[1]{equation (\ref{#1})}

\newcommand{\sub}[1]{\ensuremath{_{\mbox{\scriptsize#1}}}}

\renewcommand{\micron}{\ensuremath{\:\mu\mathrm{m}}}

\begin{document}
\slugcomment{To appear in The Astrophysical Journal}
\title{Gaps in Protoplanetary Disks as Signatures of Planets:
II.~Inclined Disks}
\author{Hannah Jang-Condell}
\affil{Department of Physics \& Astronomy, University of Wyoming, 
Laramie, WY 82071, U.S.A.}
\author{Neal J.~Turner}
\affil{Jet Propulsion Laboratory, California Institute of Technology, 
Pasadena, CA 91109, U.S.A.}
\affil{Max Planck Institute for Astronomy, 
K\"onigstuhl 17, 69117 Heidelberg, Germany}

\begin{abstract}
We examine the observational appearance of partial gaps being opened 
by planets in protoplanetary disks, considering the
effects of the inclination relative to the line of sight.
We model the disks with static
$\alpha$-models with detailed radiative transfer, parametrizing the shape
and size of the partially cleared gaps based on the results of
hydrodynamic simulations.  
As in previous work, starlight falling across the gap leads to high
surface brightness contrasts.  The gap's trough is darkened by both
shadowing and cooling, relative to the uninterrupted disk.  The gap's
outer wall is brightened by direct illumination and also by heating,
which puffs it up so that it intercepts more starlight.
In this paper, we examine the effects of inclination on resolved
images of disks with and without gaps at a wide range of wavelengths.
The scattering surface's offset from the disk midplane creates a
brightness asymmetry along the axis of inclination, making the disk's
near side appear brighter than the far side in scattered light.
Finite disk thickness also causes the projected distances of 
equidistant points on the disk surface to be smaller on the near 
side of the disk as compared to the far side.  
Consequently, the gap shoulder on the 
near side of the disk should appear brighter and closer to the star 
than on the far side.  
However, if the angular resolution of the
observation is coarser than the width of the brightened gap shoulder,
then the gap shoulder on the far side may appear brighter because 
of its larger apparent size.  
We present a formula to recover the scale height and inclination angle
of an imaged disk using simple geometric arguments and measuring 
disk asymmetries.  Resolved images of circumstellar disks have
revealed clearings and gaps, such as the transitional disk in LkCa 15.  
Models created using our synthetic imaging attempting to 
match the morphology of observed scattered light 
images of LkCa 15
indicate that the H-band flux deficit in the inner $\sim0.5\arcsec$ 
of the disk can be 
explained with a planet of mass greater than 0.5 Jupiter mass.  
\end{abstract}


\section{Introduction}

An ever-increasing number of young protoplanetary disks are being imaged, 
from space and from the ground, and across wavelengths from 
optical to infrared to radio.  These gas-rich disks are particularly 
interesting because they are the right age for giant planet formation 
to occur.  It can be tempting to interpret structures seen in these disks 
as being signatures of planet formation, but without good models including
radiative transfer, drawing these conclusions can be fraught.  
For example, scattered light images trace only the optical thin 
and diffuse layers of the disk and not the overall structure of the disk 
\citep{2007HJCBoss}.  

This paper is the second in a series of papers analyzing gap-opening 
by forming planets in disks.  
In the first paper \citep[][henceforth Paper I]{HJCTurner}, 
we carried out detailed calculations of the vertical disk temperature 
and density structure in the presence of a partial gap, 
where the density does not approach zero in the gap.  
Paper I also showed simulated images of face-on disks with partial gaps 
in both scattered light and thermal continuum images from the 
optical to the radio. 
The utility of these models is that if a 
gap is well resolved so that its depth and width can be determined, 
then we can estimate the mass of the planet opening that gap.  

Several groups have studied hydrodynamic simulations of 
planets opening gaps in disks in the absence of radiative 
heating from the central star 
\citep[e.g.][]{2008PaardekooperPapaloizou,2009AyliffeBate,2006deValBorro_etal,bate,2008MNRAS.387..387E}.  
However, illumination on these gaps can significantly 
affect both the disk structure and its observable properties 
\citep[e.g.][Paper I]{2006Varniere_etal,2012Turner_etal}. 
In particular, these studies consistently find that 
the far edge of a gap in a flared T Tauri disk 
is heated and puffed up, creating a positive feedback loop 
that enhances the appearance of the gap.  

Radiative transfer simulations on the output of hydrodynamic
simulations of disks with embedded planets
have been done \citep{2007HJCBoss}, 
some with particular emphasis toward predictions for ALMA
\citep{2005WolfDAngelo,2013Ruge_etal}.  However, incorporating 
the radiative heating from stellar illumination back into the 
hydrodynamic simulations has thus far proven to be computationally 
prohibitive.  

We model a gap or partial gap in the disk as a fixed perturbation on the 
radial surface density profile, $\Sigma(r)$ 
and carry out detailed radiative transfer on this disk structure, 
under the assumption of hydrostatic equilibrium to allow 
the disk to vertically expand or contract in response to 
heating or cooling from stellar illumination or shadowing.  
We assume an axisymmetric disk, so our models preclude 
spiral density waves.  
While a full three-dimensional hydrodynamic simulation 
including radiative transfer would be the best model for 
a gap in a disk, it is also very computationally intensive.  
The models presented here provide a useful analysis of gap opening 
in disks, even if some details are missing.  

We make use of the radiative transfer models presented in 
Paper I, which exploits analytic approximations of solutions 
to the equations of radiative transfer to efficiently 
calculate disk structure.  Monte Carlo methods are more 
accurate in a sense, since they directly track photon 
packets in the disk 
\citep[e.g.~][]{2004MNRAS.351..607W,2004DullemondDominik,2008ApJ...689..513T,2008Pinte_etal,2010MuldersDominikMin}.
However, their computationally intensive natures makes the 
iterative calculation of disk structure 
time consuming.  The methods we adopt allow for rapid iterative 
calculation of disk structure in order to attain self-consistency
between density and temperature in the structure of the disk. 
We use Monte Carlo calculations to validate selected cases.  

As an example of the application of the models presented here, 
we consider the case of the LkCa 15 pre-transitional disk.  
LkCa~15 is just one of a number of such
transitional disks whose cavities were resolved recently, including
by \citet{2012arXiv1208.2075H}, 
\citet{2012Mathews_etal}, 
\citet{2012Cieza_etal}, 
\citet{2011Andrews_etal}.
Companion planets or low-mass stars are
observed in some cases 
\citep{2012Biller_etal,2012KrausIreland,2011Huelamo_etal}.

In this paper, we examine the effects of inclination on disk 
images, making use of the models presented in Paper I\@.  
We show that some brightness asymmetries in disks may be 
purely geometrical effects, attributable to the shape of the 
last scattering surface and its changing appearance with inclination 
angle rather than anisotropic scattering.  
We show that these effects may be seen in disks both with and without 
partially cleared gaps in them.  
Finally, we apply our approach
to images of the disk surrounding LkCa 15, to constrain the
mass of the planet capable of opening the gap observed at
near-infrared and millimeter wavelengths.

\section{Model}
\label{sec:model}

The model for the disk structure and observable properties 
is the JC model presented in detail in Paper I\@.  
The disk structure model calculates radiative transfer 
based on the methods of \citet{paper1} and \citet{paper2} 
and adapted to iteratively calculate the temperature and 
density structure in a self-consistent manner 
as described in \citet{HJC_model}.  The observable properties 
are calculated following \citet{2009HJC} with modifications 
to approximate the effects of multiple scattering as 
described in Paper I.  

Here, we summarize the JC model for calculating 
disk temperature structure and synthetic images.  
The JC method is a fully three-dimensional radiative transfer 
calculation, allowing the propagation of radiation from the 
disk surface to all points within the disk, both vertically 
and radially.  It relies on a one-dimensional approximate solution 
for the equations of radiative transfer to calculate the 
propagation of photons from a given point to another, but 
does not otherwise assume a plane-parallel disk.  

The only difference between the model disks in this work 
and Paper I is a minor modification to the opacities.  
The algorithm for calculating the thermal structure of the disk 
makes use of mean opacities.  However, the mean opacities used 
in Paper I were inconsistent with the wavelength-dependent 
opacities used to calculate observables.  In this work, we have 
rectified this inconsistency by using the correct mean opacities 
and recalculating the disk structure model.  
The opacities are an input parameter to the JC code, so the 
algorithms for calculating the model disk are unchanged, although the 
specific disk structure is modified from that presented in Paper I.  
We discuss this correction to the opacities is in detail 
in \S\ref{sec:opacities}. 

\begin{figure}
\plotone{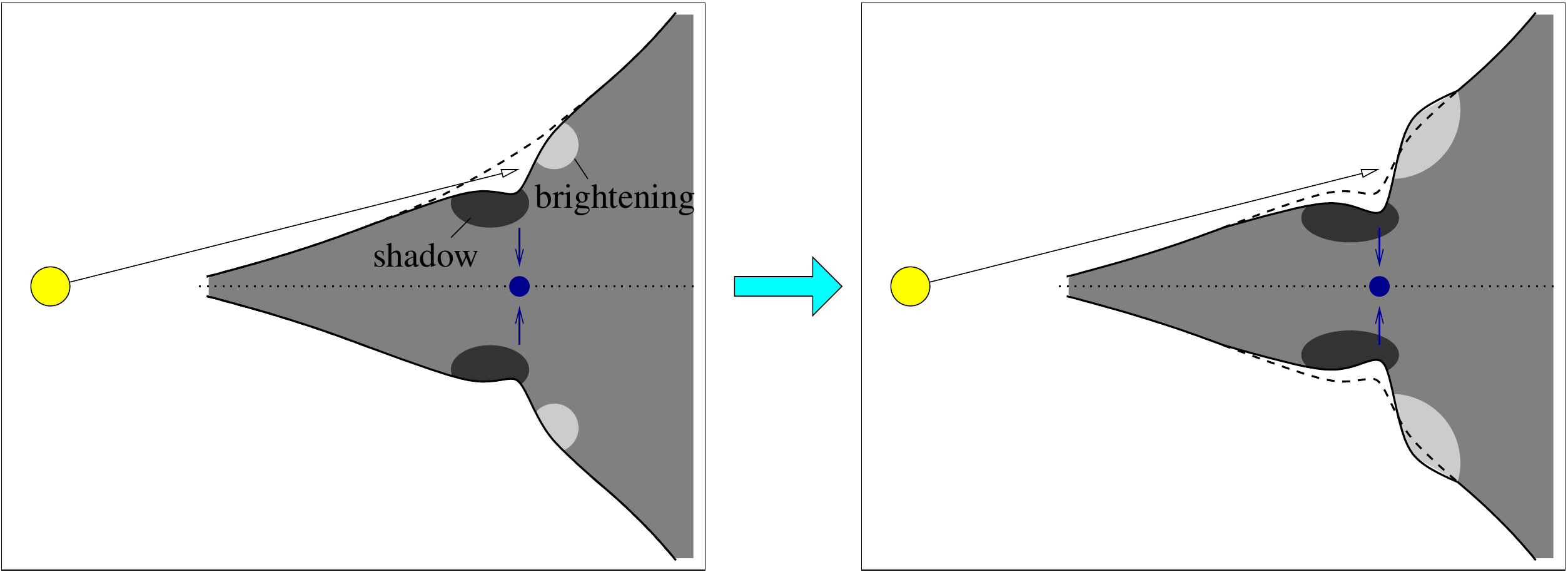}
\caption{\label{gapshadow}
Cartoon diagram of radiative feedback on disk structure.  
The star is represented as a yellow disk, and the planet by a blue dot.  
The disk surface represents a contour of roughly constant density.
The left image shows the initial gap opened in the disk, 
with the dotted line showing the original, unperturbed disk 
surface.  Stellar illumination on the surface of the gap 
creates shadowed and brightened regions.  
Shadowing and cooling occurs in the disk trough, and the far 
side of the gap is illuminated and heated.  
The right image shows the response of the gaseous disk material to 
the cooling and heating: the shadowed region contracts and deepens the 
gap, while the illuminated far side expands and is elevated.  
}
\end{figure}

Gap opening is modeled as a Gaussian-shaped 
ad hoc perturbation to the surface density profile.  
We consider partially cleared gaps, where the density 
does not go all the way to zero within the gap.  
The disk temperature is then 
iteratively recalculated to account for the thermal adjustment 
of the disk to shadowing in the gap and illumination 
on the exposed far gap wall (see \figref{gapshadow}).

Radiative heating of the disk is integrated piecewise over the surface 
of the disk, allowing for three-dimensional propagation 
of radiation throughout the disk.  
The parameters used to model the fiducial disk are listed in 
Table \ref{table:params}.
The stellar parameters are 
mass $M_*=1\,M_{\odot}$, 
radius $R_*=2.6\,R_{\odot}$, 
and effective temperature $T\sub{eff}=4280$ K, 
consistent with a protostar with an age of 1 Myr \citep{siess_etal}.  
The luminosity of a star with these parameters is 
$L_* = 4\sigma_B\pi R_*^2 T\sub{eff}^4 = 2.04 L_{\odot},$
where $\sigma_B$ is the Stefan-Boltzmann constant.  
We assume a constant-$\alpha$ disk model where the 
viscosity is parameterized as $\nu=\alpha\sub{ss} c_s H$
\citep{shaksun}, where 
$c_s(r)$ is the sound speed at the midplane, and $H(r)$ is the thermal scale 
height of the disk at radius $r$ from the star.  
The disk parameters are 
accretion rate $\dot{M}=10^{-8}\,M_{\sun}\,\mbox{yr}^{-1}$ and 
viscosity parameter $\alpha\sub{ss}=0.01$, 
parameters typical for T Tauri stars.  

\begin{deluxetable}{ccc}
\tablecaption{\label{table:params}Stellar and disk parameters used for disk models}
\tablehead{
\colhead{parameter} & 
\colhead{fiducial disk} & 
\colhead{LkCa 15}} 
\tablecolumns{3}
\startdata
$M_{*}$ ($M_{\odot}$) & 1 & 1
\\
$R_{*}$ ($R_{\odot}$) & 2.6 & 1.51 
\\
$T\sub{eff}$ (K) & 4280 & 4350
\\
$L_{*}$ ($L_{\odot}$) & 2.04 & 0.74
\\
$\alpha\sub{SS}$ & 0.01 & 0.0007 
\\
$\dot{M}$ ($M_{\odot}\,\mbox{yr}^{-1}$) & $10^{-8}$ & $2.4\times10^{-9}$
\enddata
\end{deluxetable}

\subsection{Opacities}
\label{sec:opacities}

\begin{figure}
\plotone{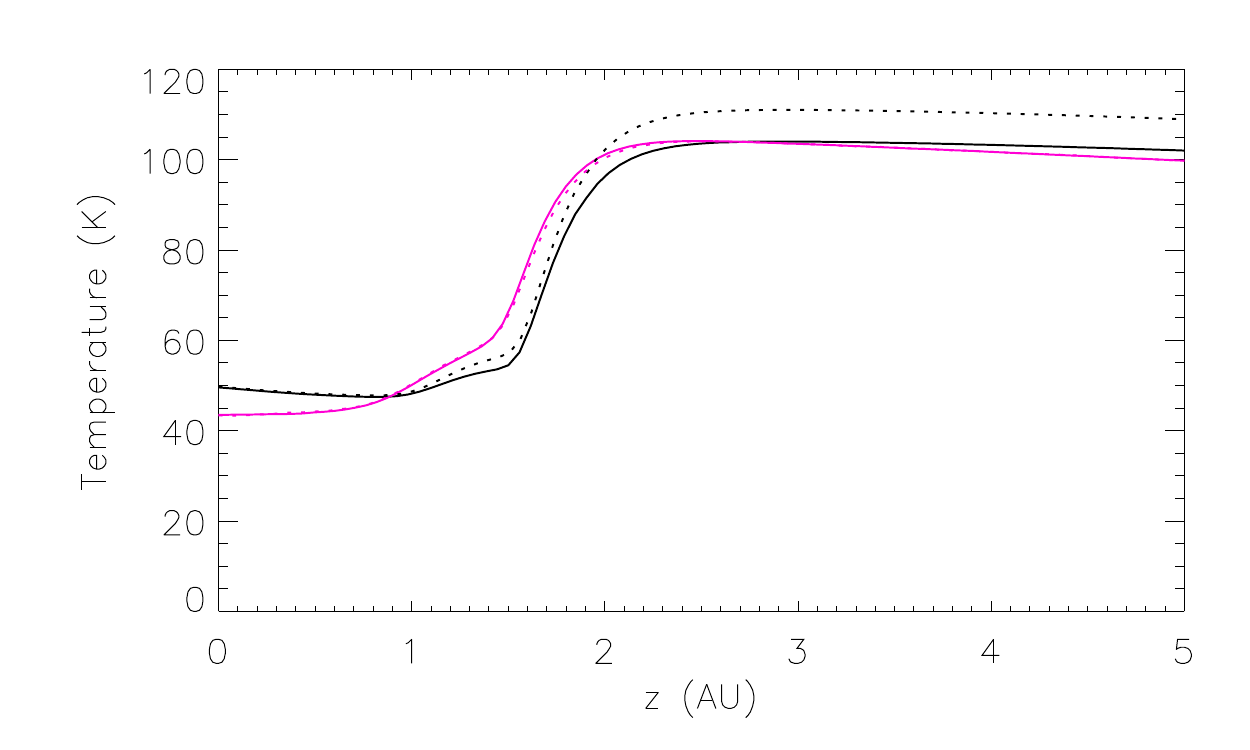}
\caption{\label{verttemp}
Vertical temperature profile of the unperturbed disk at 10 AU.  
JC model calculations are shown in black, and Monte Carlo 
comparison models are shown in magenta.  Dotted lines 
show results from Paper I, and solid lines are this work.  
Incorrect mean opacities were implemented in 
the JC models in Paper I, yielding 
disk surface temperatures that were too high. 
Using the correct opacities, 
the solid black and magenta lines are nearly coincident
above 2 AU\@.
}
\end{figure}

The mean opacities, consistent with the wavelength-dependent 
opacities used for calculating observables, are as follows:
the Rosseland mean opacity of the disk is $\chi_R = 2.72 \mbox{ cm g}^{-2}$, 
the disk temperature averaged absorption is 
$\kappa_P = 1.45 \mbox{ cm g}^{-2}$, 
and the mean absorption and extinction averaged over the stellar spectrum 
are $\kappa_P^* = 1.33 \mbox{ cm g}^{-2}$ and 
$\chi_P^* = 10.88 \mbox{ cm g}^{-2}$, respectively.  
In Paper I, the opacities adopted were 
$\chi_R = 2.61, 
\kappa_P = 1.35,
\kappa_P^* = 1.61$,
and 
$\chi_P^* = 11.75
\mbox{ cm g}^{-2}$.
The incorrect mean opacities resulted in temperatures at 
or above the disk surface to be too high compared 
to the Monte Carlo (MC) models, as shown in 
\figref{verttemp}.  The effect on the surface temperature can be 
explained as follows.  As described in 
\citep{HJC_model}, the temperature can be approximated as 
\begin{equation}\label{irradflux}
\frac{\sigma_B T_r^4}{\pi} = \frac{\kappa_P^*}{\chi_P^*}
\frac{F\sub{irr} \mu}{4\pi}
   \left[ c_1 + c_2 e^{-\tau_s} + c_3 e^{-g \mu\tau_s} 
     \right], 
\end{equation}
where 
$\mu(r)$ is the cosine of the 
angle of incidence of stellar irradiation at the surface 
of the disk, 
$F\sub{irr}(r)$ is the stellar flux at radius $r$, 
$g=\sqrt{3(1-\omega)}$, 
and $\tau_s$ is the optical depth to the star.  
The coefficients are 
\begin{eqnarray}
c_1 &=& \frac{ 6 + 9\mu\chi_R/\chi_P^* }{g^2}
- \frac{6 (1-\chi_R/\chi_P^*)\left(3-g^2\right)}{g^2
   (3 + 2 g) (1 + g \mu)} \label{c1}
\\
c_2 &=& 
\left(\frac{\chi_P^*}{\mu \kappa_P}-\frac{3\mu\chi_R}{\chi_P^*}\right)
\frac{(1-3\mu^2)}{(1-g^2 \mu^2)} \label{c2}
\\
c_3 &=& \left(\frac{g \chi_P^*}{\kappa_P}-\frac{3\chi_R}{\chi_P^*g}\right)
\frac{(2+3\mu)(3-g^2)}{g(3+2g)(1-g^2 \mu^2) }. \label{c3}
\end{eqnarray}
Above the surface, $\tau\rightarrow0$ and $\mu\rightarrow0$.  
Then, to leading order in $\mu$, 
$c_1 + c_2 + c_3 \rightarrow \chi_P^*/\mu \kappa_P$ and 
\begin{equation}
T_r^4 = \frac{\kappa_P^*}{\kappa_P}\frac{F\sub{irr}}{4 \sigma_B}
\end{equation}
The ratio $\kappa_P^*/\kappa_P$ is decreased from 1.20 to 0.92, 
resulting in a temperature decrease at low optical depths of 
about 6\%, which is the temperature decrease observed 
between Paper I and the fiducial disk model in this work at 
a radius of 10 AU and height 2.5 AU, shown in 
\figref{verttemp}.

Although the temperature difference is $\sim10-20$ degrees, 
the alteration to the temperature occurs in the upper 
layers of the disk, several times the thermal scale height 
of the disk, which is $H=0.48$ AU at $r=10$ AU\@.  
The density of the disk is quite low at these heights.  
On the other hand, the temperature at the midplane, the densest 
region, is largely unchanged.  Thus, the overall structure 
of the disk changes only slightly between Paper I and this work.

\subsection{Disk Structure}

As described in Paper I and references therein, the density and
temperature structure of the disk models 
are calculated iteratively, keeping
the total vertically-integrated surface density fixed.  
We refer to the disk without a gap as the fiducial model.  
Disks with gaps are calculated in the same way as the fiducial 
model, the difference being in the surface density profile used.  
The heating of the disk from stellar irradiation is calculated by piecewise
integration over the disk surface.  The density profile of the disk is
then recalculated under the assumption of hydrostatic equilibrium.
This process is iterated until the disk structure converges.  We
assume azimuthal symmetry for disk models both with and without gaps,
but accounting for the three-dimensional curvature of the disk.  The
surface density of the gapless disk is given as $\Sigma_0$, and a disk
with a gap as $\Sigma$ without a subscript.

\subsection{Gap Opening}

\begin{figure}
\plotone{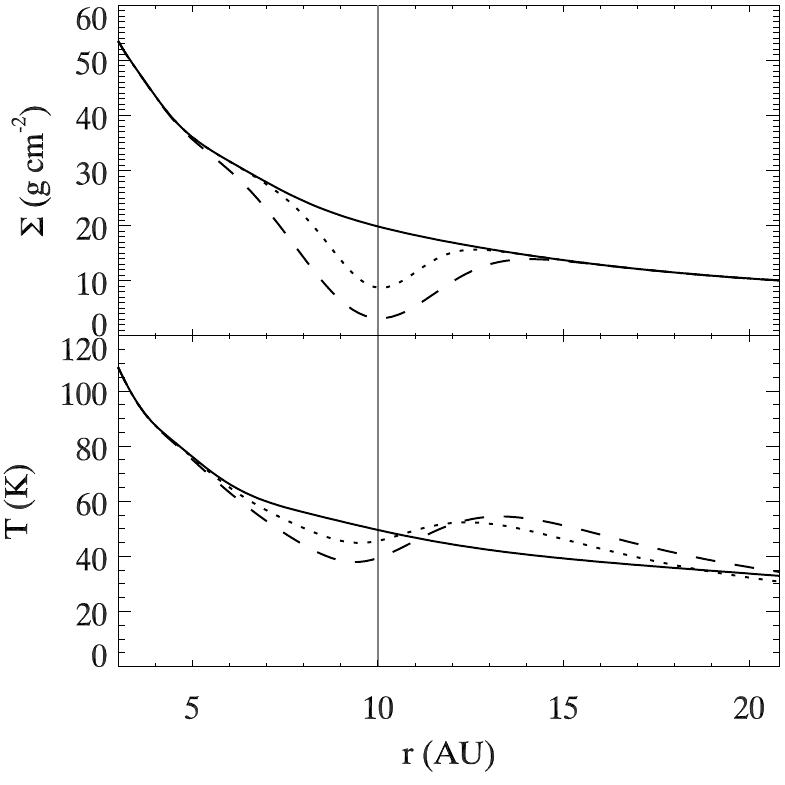}
\caption{\label{dentemp}
Surface density (top) and midplane temperature (bottom) profiles 
for a disk with and without a gap created by a planet at 10 AU\@. 
Solid, dotted, and dashed lines indicate 
planet masses of 0, 70, and 200 $M_{\oplus}$, respectively.  
The thermal perturbation is caused by shadowing and illumination
by stellar irradiation at the disk surface.  
}
\end{figure}

Gap opening by planets is modeled as an ad hoc modification to the surface 
density profile so that 
the surface density of a disk modified by a gap 
of width $w$, depth $d$, and position $a$ is
\begin{equation}\label{eq:gap}
\label{gapdenprof}
\Sigma(r) = \Sigma_0(r) \left\{1-d\exp[-(r-a)^2/(2w^2)]\right\}.
\end{equation}
The values of $d$ and $w$ versus planet mass are determined 
from the results of \citet{bate}, using the gap-opening 
criterion of \citet{2006CridaMorbidelliMasset}, 
\begin{equation}\label{eq:viscgapcrit}
G \equiv \frac{3}{4}\frac{H}{r\sub{Hill}} + \frac{50}{q {\cal R}}
\lesssim 1
\end{equation}
where we have defined $G$ to be the gap-opening parameter, 
$r\sub{Hill}$ is the Hill radius, 
$q$ is the ratio of planet to stellar mass, and 
the Reynolds number is ${\cal R}\equiv r^2\Omega_P/\nu$.
In the models analyzed in this paper, a 70 (200) $M_{\oplus}$ 
planet at 10 AU opens a gap of $d=0.56 (0.84)$ and 
width $w=1.1 (1.7)$ AU, as listed in Table \ref{fittable}.
The shapes of the gaps are illustrated in \figref{dentemp}.
We refer the gaps created by planets with $G<1$ to be 
{\em partial} gaps because the density 
does not go to zero or nearly zero within the gap.  

\begin{deluxetable*}{ccccc}
\tablecaption{\label{fittable}Best-fit parameters for 
gaps opened by planets.}
\tablehead{
\colhead{mass ratio\tablenotemark{1} ($q$)} &
\colhead{$3\times10^{-5}$} & 
\colhead{$1\times10^{-4}$} & 
\colhead{$3\times10^{-4}$} & 
\colhead{$1\times10^{-3}$} 
}
\startdata
$d$ &
0.014           & 0.56             & 0.84             & 0.99
\\ 
$w/a$ &  
0.078           & 0.11             & 0.17     &---\tablenotemark{2}
\\
gap-opening parameter ($G$) & 
18.4            & 6.2              & 2.5              & 1.04 
\\ \cutinhead{Fiducial Disk Model (planet at 10 AU, $H/a=0.048$)}
derived $q$ &
$6.7\times10^{-5}$ & $2.2\times10^{-4}$ & $6.2\times10^{-4}$ & $1.9\times10^{-3}$
\\
derived planet mass\tablenotemark{3} & 
22 $M_{\earth}$ & 72 $M_{\earth}$ & 210 $M_{\earth}$ & 620 $M_{\earth}$
\\ \cutinhead{LkCa 15 Model}
planet position\tablenotemark{4} ($a$) & 
\nodata &  40.7 AU & 38.3 AU & 32.5 AU \\
disk aspect ratio\tablenotemark{5} ($H/a$) & 
\nodata & 0.0609 & 0.0600 & 0.0569 \\
derived $q$ & 
\nodata &
$3.1\times10^{-5}$ & $1.08\times10^{-4}$ & $4.6\times10^{-4}$ \\
derived planet mass &  
\nodata & 10.6 $M_{\earth}$ &  37.3 $M_{\earth}$ & 154 $M_{\earth}$
\enddata
\tablenotetext{1}{As simulated in \citet{bate}}
\tablenotetext{2}{The gap opened by the $q=10^{-3}$ planet 
in the \citet{bate} simulation is not well-modeled by a Gaussian.}
\tablenotetext{3}{Actual masses used for this work.  In the text,
the masses have been rounded to 20, 70, and 200 $M_{\oplus}$
for convenience.}
\tablenotetext{4}{Determined by wall creation at 46 AU}
\tablenotetext{5}{Calculated from disk properties at planet position}
\end{deluxetable*}

Stellar illumination on the the gap creates a shadow in the 
gap trough and increased illumination on the exposed 
far gap wall.  The shadowed trough cools and contracts 
while the far gap shoulder heats and expands, 
changing the vertical density structure of the gap, 
as illustrated in \figref{gapshadow}.  
The density and temperature structure of the gapped disks are 
iteratively recalculated to account for these changes.  
The resulting midplane temperature profiles are 
plotted in \figref{dentemp}.

\subsection{Observables}
\label{sec:observables}

We consider two main modes of disk emission: scattered stellar light and
thermal emission from the disk.  Full details of the calculations are 
described in Paper I, but we summarize the essential model below. 

Scattered light contributions are calculated using local conditions in
the surface of the disk and the geometry of the scattering angles.
The scattering surface is determined to be where the optical depth to
light from the star at frequency $\nu$ is $\tau_{\nu,*}=2/3$. In the case of
single isotropic scattering, the brightness of scattered light from a
given point on the scattering surface is
\begin{equation}\label{eq:sglscat}
I_1^s(\nu)=\frac{\omega_{\nu}\mu\/R_*^2\/B_{\nu}(T_*)}{4r^2(\mu+\cos \eta)},
\end{equation}
where $\mu$ is the cosine of the angle of incidence of stellar light on
the surface, $\omega_{\nu}$ is the wavelength-dependent albedo, $r$ is
the distance between the surface and the star, $\eta$ is the viewing 
angle, defined as the angle of
scattering to the observer with respect to the surface normal,
and $B_{\nu}(T_*)$ is the stellar
brightness at $\nu$, evaluated as the Planck function.

The contribution to disk brightness from photons scattered two 
or more times is 
\begin{eqnarray}\label{eq:multscat}
I_2^s &=&   
\frac{B_{\nu}(T_*) R_*^2}{4 r^2}
  \frac{\mu \omega^2}{1-g^2\mu^2} 
\times \\ & &
    \left[ \frac{2+3\mu}{(1+2g/3)}
    \frac{1}{(1+g\cos\eta)}
      - \frac{3\mu}{(1+\cos\eta/\mu)}
\right]
\nonumber
\end{eqnarray}
where $g=\sqrt{3(1-\omega_{\nu})}$ and isotropy is assumed.  
Then the total brightness from scattered light is 
\begin{eqnarray}\label{eq:totscat}
I_{\nu}^s &=& I_1^s + I_2^s =
\frac{\omega_{\nu}\mu\/R_*^2\/B_{\nu}(T_*)}{4r^2(\mu+\cos \eta)}
\times \\ &&
\left\{
1 + \frac{\omega_{\nu}}{1-g^2\mu^2} 
  \left[\frac{(2+3\mu)(\mu+\cos\eta)}{(1+2g/3)(1+g\cos\eta)} - 3\mu^2
\right]
\right\}
\nonumber
\end{eqnarray}

\begin{figure}[htbp]
\plotone{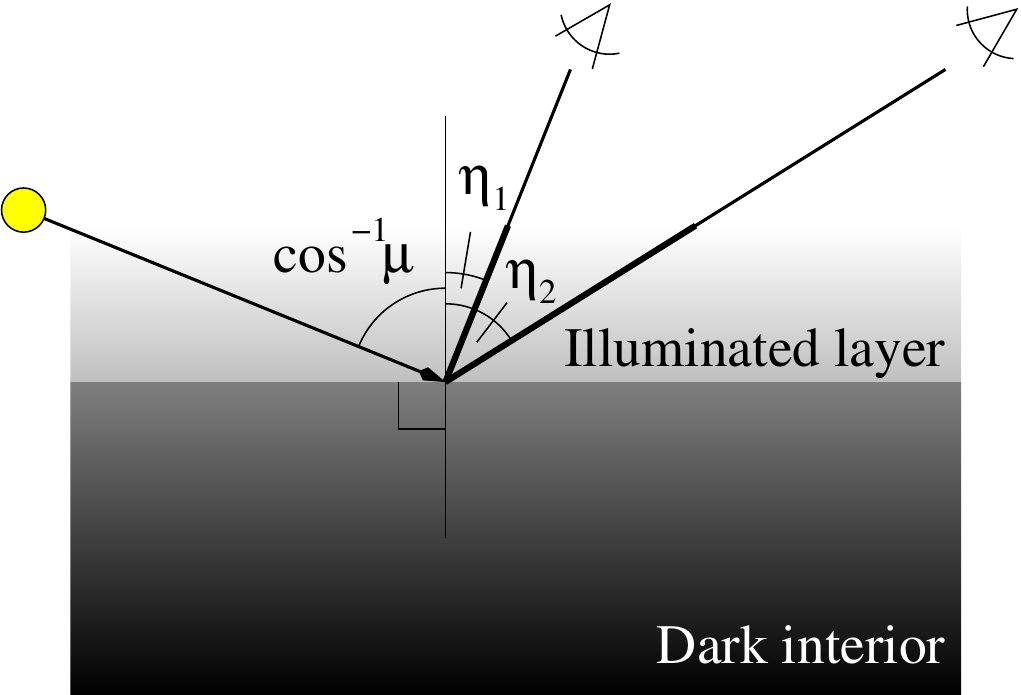}
\caption{\label{schematic}Illustration of how inclination affects 
brightness of scattered light.  
The path length through the illuminated layer represents the relative 
brightness seen by each observer. 
The cosine of the angle of incidence of stellar light at the surface 
is $\mu$.  The angle between the surface normal and the observer is $\eta$.  
For $\eta_2>\eta_1$, there are more scatterers along the path 
through the illuminated layer for the observer at $\eta_2$, so the 
disk appears brighter.  
}
\end{figure}

Note that the viewing angle $\eta$ is distinct from the inclination 
angle $i$, as $i$ is a global value assigned to a disk referring to 
the tilt of the disk midplane with respect to the observer, whereas 
$\eta$ is a local value within the disk, and depends on both $i$ and 
the local disk shape.  
The angles are illustrated in \figref{schematic}, 
which shows that for larger values of $\eta$ ($\eta_2>\eta_1$), 
the observer sees 
more scatterers along the line of sight, resulting in an 
apparent brightening of the surface layer.  

The thermal emission is calculated as in Paper 1 to 
include multiple scattering of thermally emitted photons within the 
disk.  The source function is given by 
\begin{equation}\label{source}
S_{\nu} = \omega_{\nu} J_{\nu} + (1-\omega_{\nu}) B_{\nu}
\end{equation}
where $\omega_{\nu}$ is the wavelength-dependent albedo, 
$B_{\nu}=B_{\nu}(T)$ is the local thermal emission, and 
$J_{\nu}$ is the mean intensity.  
The mean intensity, assuming local thermodynamic equilibrium, 
is given by 
\begin{equation}\label{Jisotherm}
J_{\nu} = B_{\nu}(T) \left[
1 
- \frac{\exp\left[g(\tau_{\nu}-\tau\sub{max})\right]
+\exp\left(-g\tau_{\nu}\right)
}{
\left(1-g/\sqrt{3}\right)
\exp\left(-g\tau\sub{max}\right)
+1+g/\sqrt{3}
}
\right],
\end{equation}
where $\tau_{\nu}/\cos i$ is the optical depth at frequency $\nu$ integrated along the line of sight and $\tau\sub{max}$ is the maximum value of $\tau_{\nu}$ when integrated all the way through the disk.
The emitted intensity $I_{\nu}^t$ is found by 
integrating the equation of radiative transfer for a locally 
plane-parallel medium,  
\begin{equation}\label{intensity}
\cos\eta \frac{\partial I_{\nu}^t}{\partial\tau_{\nu}} = I_{\nu}^t-
\omega_{\nu} J_{\nu} - (1-\omega_{\nu}) B_{\nu}
\end{equation}
with the assumption that 
the disk is geometrically thin ($\eta \approx i$).  
The mean intensity is evaluated along the 
line of sight through the disk, and 
$\tau_{\nu}$ is measured as the optical depth 
perpendicular to the disk.

A disk image is assembled pixel by pixel.  Each pixel represents 
a particular line of sight through the disk.  
For scattered light images, we find where the line of sight intercepts
the disk surface, and calculate $r$, $\mu$ and $\eta$ at that point to
obtain the disk brightness from Equation (\ref{eq:totscat}).
For thermal emission, Equation (\ref{intensity}) is integrated along
the line of sight represented by each pixel.

\section{Results}

Using the methods outlined in \S\ref{sec:model} and 
detailed in Paper I, we calculate the thermal
structure of a protoplanetary disk with and without gaps 
created by planets.  
We choose two gap sizes, corresponding 
the planet masses of 70 and 200 $M_{\earth}$.  
These models are the same as those presented in Paper I, 
with a correction to the opacities that produces only a minor 
modification to the disk structure in the low density upper layers.  
We then simulate images of these disks at a range 
of wavelengths ($1, 10, 30, 100, 300,$ and $1000\micron$), 
and varying inclination angle ($0\degr, 30\degr, 45\degr$, and $60\degr$.

\subsection{Validation}

As in Paper I, we compare the resulting disk models to 
Monte-Carlo (MC) models \citep{2012Turner_etal}.  
Where not otherwise specified, our MC runs use $10^9$ photons each.  
The final two-dimensional density structure (assuming azimuthal symmetry) 
of each disk model is fed into the MC code to produce a comparison 
temperature model of the disk, and accompanying observables.  

In Paper I, we demonstrated that the JC and MC models give 
qualitatively consistent results for the thermal structure of the disk.  
One inconsistency was that the temperatures above the disk surface 
were hotter in the JC models than the MC models.  
This was a result of incorrect mean opacities used as 
input parameters, and this has been corrected in the present 
work and explained in detail in \$\ref{sec:opacities}.  
The JC models presented here are otherwise 
calculated in exactly the same way as in Paper I.  
Since the temperatures affected were in the upper, least dense 
layers of the disk, the effect of this correction on the 
overall disk structure was minor, and the results of 
Paper I are not significantly affected.  

\begin{figure*}[htbp]
\includegraphics[width=0.5\textwidth]{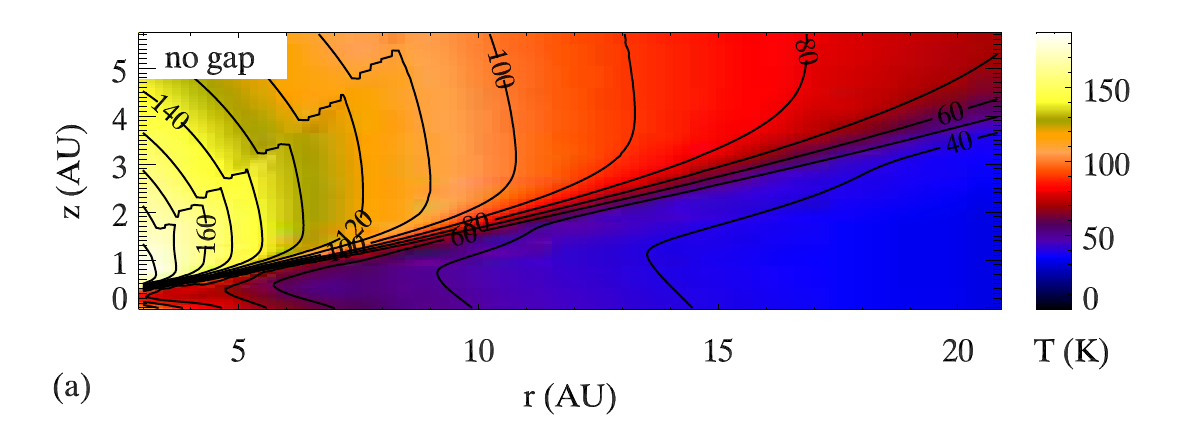}
\includegraphics[width=0.5\textwidth]{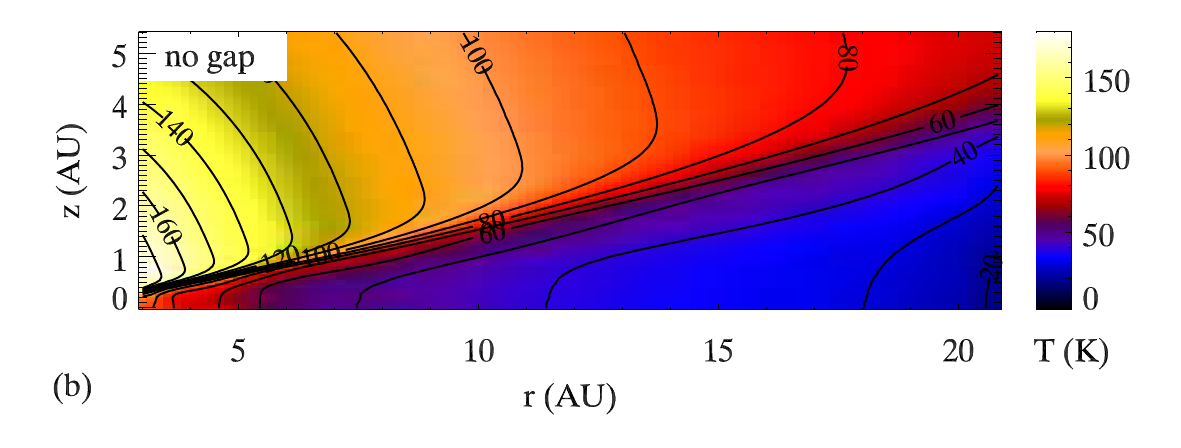}\\
\includegraphics[width=0.5\textwidth]{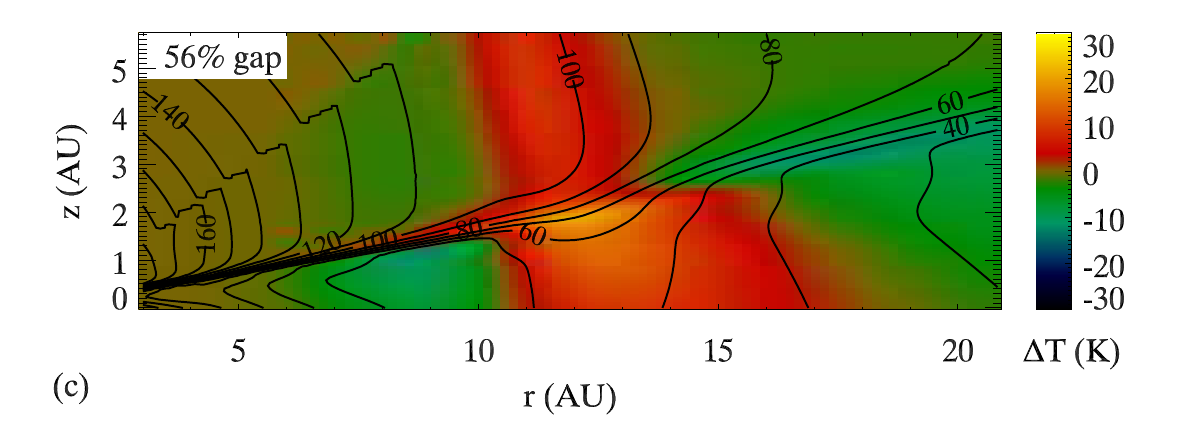}
\includegraphics[width=0.5\textwidth]{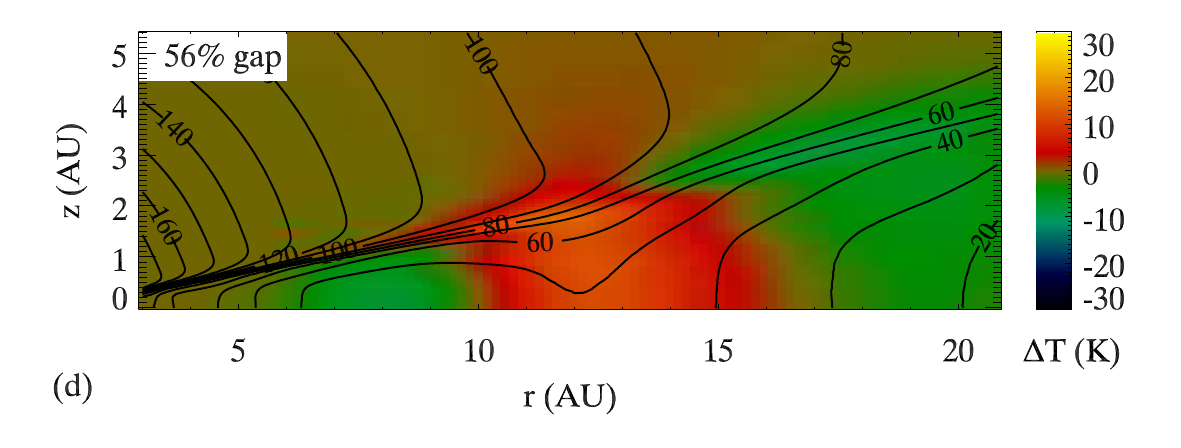}\\
\includegraphics[width=0.5\textwidth]{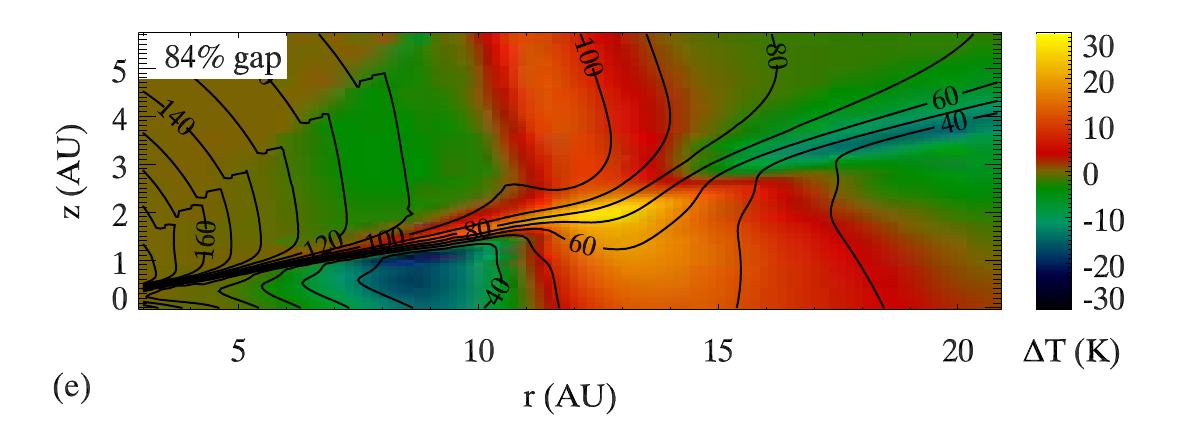}
\includegraphics[width=0.5\textwidth]{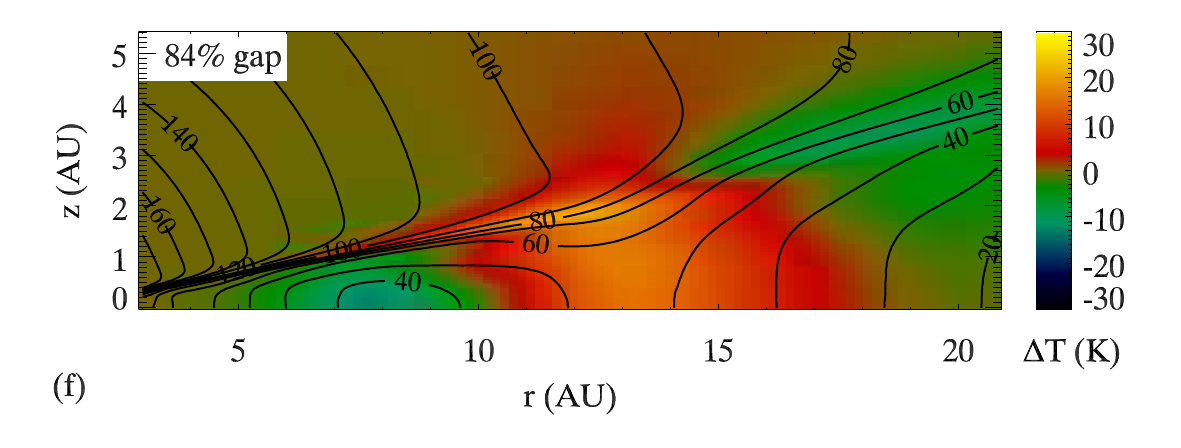}\\
\caption{\label{tempslices}Temperature cross-sections of 
model disks, where $r$ is the radial distance 
and $z$ is vertical height above the midplane.  
The JC model is shown in subfigures (a), (c), and (e), 
while the MC model is shown in (b), (d), and (f).  
The height axis is scaled to the radial axis, so the aspect ratio is correct.  
Contours indicate temperature in Kelvins and are spaced at 10K intervals.   
(a,b): The temperature profile of a disk without a planet or gap.  
The color scale shows the temperature.  
(c-f): The temperature profiles of disks with gaps at 10 AU from the
star, and created by a 70 M$_{\oplus}$ (c,d) and 200 M$_{\oplus}$
planet (e,f).  Contours show the absolute temperature, and the color
scale shows the temperature difference ($\Delta T$) as compared to
that in the unperturbed gap-less disk, with green to blue colors
showing regions that have cooled and red to yellow colors showing
regions that have heated.  
}
\end{figure*}

In \figref{tempslices}, we show the temperature cross-sections 
as contour plots resulting from the JC (a,c,e) 
and MC (b,d,e) models.  
Figures \ref{tempslices}c--\ref{tempslices}f 
show the cooling and heating created by shadowing and 
illumination on gaps in the disk, as $\Delta T$.  
The apparent heating at the disk 
surface seen at 10 AU and inward is caused by the 
cooling and contraction of the disk, which lowers the disk surface.  
Since the disk surface is hotter than the interior, 
the region of space now above the photosphere is 
hotter that it was in the unperturbed disk case.  

As noted in Paper I, the temperature structure of the disk 
is qualitatively similar in both the JC and MC radiative transfer models.  
However, there were a few notable deviations.  
First, the temperature of the surface was $\sim$10 K hotter in the 
JC model.  This has been corrected by implementing the correct values 
for the mean opacities, as illustrated in \figref{verttemp}. 
As the temperature corrections occur mostly in the upper  
layers of the disk, where the density is low,  
the change to the structure of the disk is small.  
Thus, the results presented in Paper I are qualitatively still 
relevant.  

Second, the midplane temperatures are $\sim10 K$ hotter 
in the JC model.  This is likely a result of the fact that the analytical 
solution used to estimate heating contributions from individual surface 
elements to points within the disk assumes a semi-infinite slab.  
This would result in overheating at high optical depths, since 
in a finite-thickness medium, photons could escape out the 
other side of the slab. 

Third, the depth of surface heating is deeper in the MC model, so that 
the photosphere is generally hotter.  This is likely because 
the use of mean opacities in the JC model means that all photons are 
treated as having the same opacity, while the MC model 
implements wavelength-dependent opacities, allowing photons 
longward of the stellar blackbody peak but still shortward 
of the disk thermal peak to penetrate deeper into the disk and heat 
those regions.  This explains the additional cooling seen 
in the JC model in \figref{tempslices}e just inside 10 AU, 
as well as the closer spacing of temperature contours near the 
disk surface in the JC model.  
Correcting this would involve abandoning the 
analytic solution to the radiative transfer equation that enables 
the speed of the computation, but this is outside the scope of the 
current paper.  

Fourth, the JC models show a greater degree of heating than the MC 
models above the surface at radii just above the radial position of 
the gap, as shown by the red regions in 
Figures \ref{tempslices}c and \ref{tempslices}e 
at around $r=12$ AU and $z\gtrsim2$ AU\@.  
This is due to the interpolation of temperatures 
above and below the surface in the JC model.  
In constrast, the MC models show little change in temperature 
above the disk surface, as shown in 
Figures \ref{tempslices}d and \ref{tempslices}f.   
As this region above the disk surface is 
diffuse and contributes little to the disk structure or to observations, 
the difference in temperatures does not significantly 
affect either the interior of the disk or any observables.  
Thus, we set aside the resolution to this problem for future work.  

As discussed in Paper I, despite the differences in the 
temperature structures of the JC and MC models, 
they give good qualitative agreement.  
Examining Figures \ref{tempslices}c-f,
both models show that when a partial gap is opened in a disk, 
material is cooled in the gap trough (the green areas around 
$r=10$ AU) while the 
far wall is heated (the red to yellow regions outward 
of 10 AU).  The models also agree that the 
heated and puffed up outer gap wall should shadow the 
regions still further outward (green areas roughly around 
$r>15$ AU and $z\sim30$ AU).
This effect was also seen in models of solar illumination 
on the gap opened by Jupiter \citep{2012Turner_etal}, which 
used the same radiative transfer modeling as the MC models.  

\begin{figure*}[htbp]
\plotone{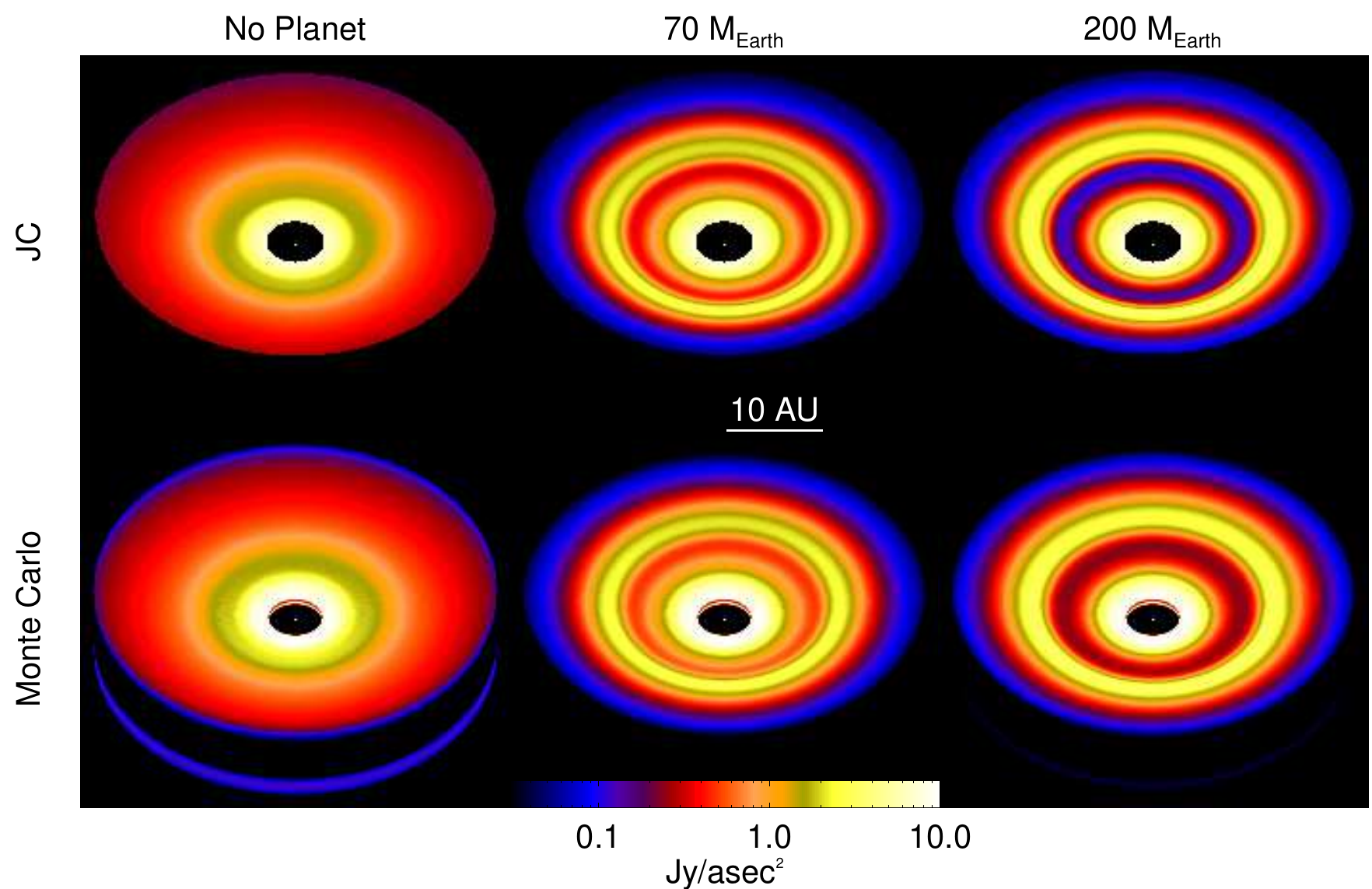}
\caption{\label{comparetilted}
Comparison of scattered light images using JC models (top) 
and a Monte Carlo code (MC, bottom).  The disks are tilted at $45\degr$ and 
the images are calculated at 1 micron.  
The left panels show disks with no gaps.  The center/right 
panels show disks with gaps opened by a 70/200 $M_{\oplus}$ 
planet.  JC models and MC models give similar results, validating 
our use of the JC models henceforth.  
}
\end{figure*}

We calculate 1 $\mu$m scattered light model images of our disks 
with both the JC and MC models with an inclination of 45\degr, 
and compare our results in \figref{comparetilted}. 
Overall, there is good agreement in the morphology of both sets 
of disk models.  The brightness of the far gap wall is similar in 
magnitude, and the shape of the bright ring is nearly identical.  

The differences in the two sets of models are the same model 
differences seen in Paper I, for face-on disks.  
As seen in Paper I, the JC models predicts a 
deeper shadow in the gap trough than the MC model.  This is likely 
because while the JC model treats multiple scattering locally, 
it does not consider photons scattered into the 
gap from the brightened far gap wall, which may brighten the gap trough.  

The MC model also predicts a brighter inner disk, especially close to the 
star.  This may be a result of a different treatment of the inner 
boundary condition.  
The JC code assumes an analytic profile for the inner disk, while the
MC calculations treat no transfer there, simply discarding each photon
arriving directly from the star that would be absorbed or scattered in
the innermost modeled annulus.

Although there are some differences between the simulated images of 
the disks, qualitatively they show the same features: 
darkening in the gap trough and brightening on the far gap shoulder.  
Moreover, the morphology of the images of tilted gaps is 
nearly identical.  In both the JC and MC models, the 
near side of the disk (bottom half of each image) appears 
forshortened in comparison to the far side.  This is evidenced 
by the narrowing of both the gap shadow and the brightened outer 
shoulder on the near side of the images.  Moreover, the near 
side of the disk is slightly brighter than the far side of the disk 
in both the MC and JC models.  The reasons for these effects 
are described in the remainder of this section.  

\subsection{Simulated Images of Inclined Disks}

\begin{figure}[htbp]
\plotone{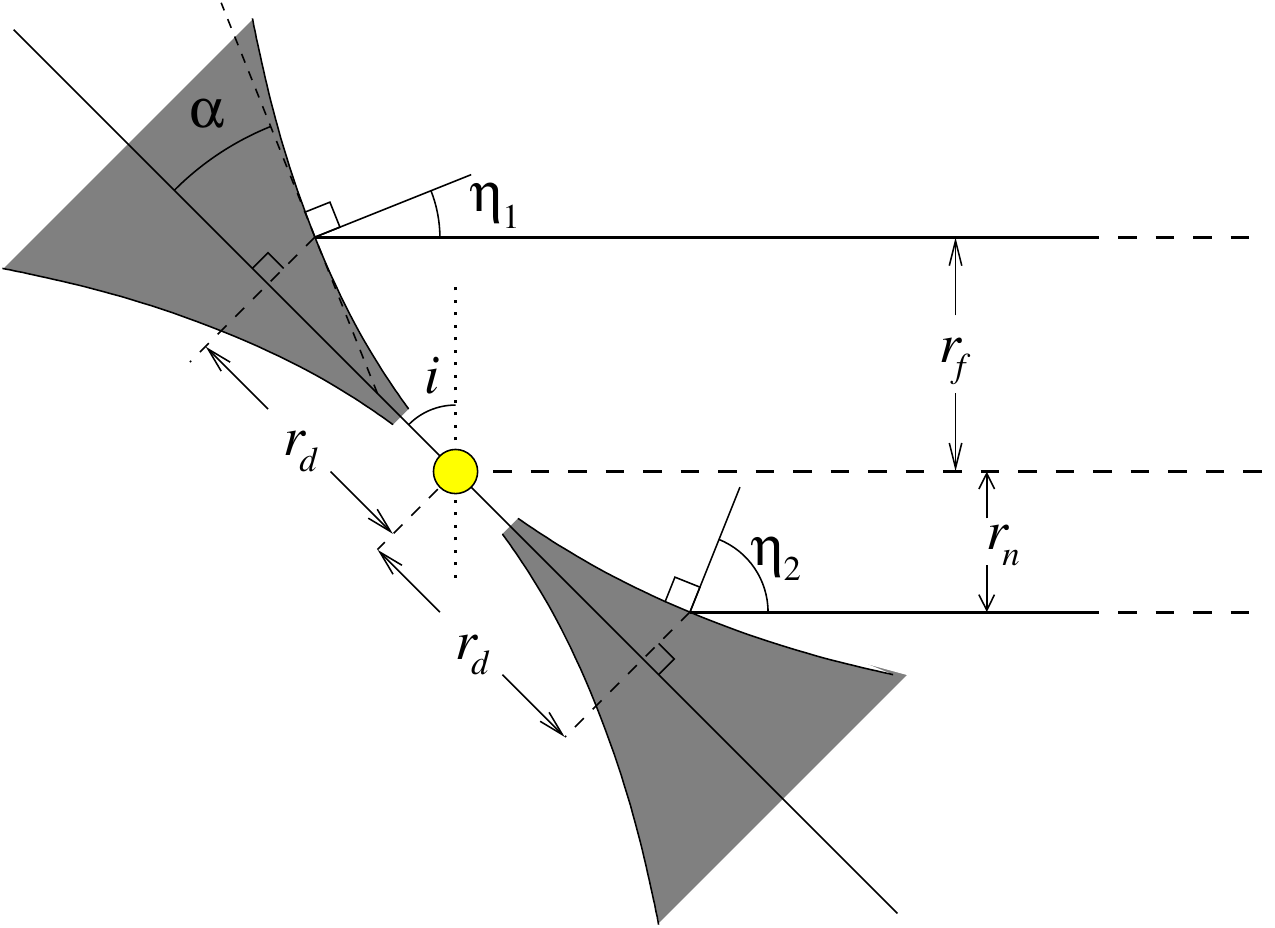}
\caption{\label{tilteddisk}
Schematic of inclination and aspect angles in the disk.  
The yellow circle represents the central star and 
the gray flared wedges represent the optically thick interior of the disk
in cross-section.  
Points that are equidistant from the star in the deprojected disk 
with a distance of $r_d$ appear to be at different 
distances on the near side versus the far side of the disk 
when the disk is inclined at an inclination angle of $i$.  
The observer sees only the surface of the disk above the 
optically thick region.  
The aspect angle $\alpha$ is the angle that this surface 
makes with the disk midplane. 
The points on the near and far sides of the 
disk appears to be at a distances $r_n$ and $r_f$, respectively.  
Moreover, the viewing angles along 
these lines of sight are different, with 
$\eta_1 = i-\alpha$ and $\eta_2 = i+\alpha$.  
Since $\eta_2>\eta_1$, the nearer line of sight traverses a 
greater path length in the bright disk atmosphere and therefore 
appears brighter than the far point.  
}
\end{figure}

Henceforth, unless specifically referred to otherwise, 
the disk models discussed are the JC models.  
We are interested in the effect of the appearance of disks 
with and without gaps as the inclination angle varies.  
To aid in the interpretation of the effects of inclination 
on disk images, we present a schematic of disk tilted 
at an inclination $i$ with respect to the observer 
in \figref{tilteddisk}, and will refer to the angles and distances 
defined in it throughout.  

In particular, we note that an axisymmetric structure of radius $r_d$ 
on the surface of the disk, when projected on the sky, will appear 
asymmetric along the minor axis, having a projected distance 
of $r_n$ on the near side of the disk (the lower right 
half of the schematic disk) and $r_f$ on the far side of the disk 
(upper left half).  This is because of the finite thickness 
of the disk: if the structure was in the midplane of the disk, 
then $r_f=r_n$.  

We also note that if the slope of the disk surface is 
given by $\tan\alpha$ at the position of this same 
hypothetical axisymmetric disk structure, then the 
local viewing angles, 
$\eta_1$ and $\eta_2$ for the far side and near side, 
respectively, are not equal, and are given by 
$\eta_1 = i-\alpha$ and $\eta_2 = i+\alpha$.  
This will become relevant in discussing brightness 
asymmetries seen in the simulated disk images.

\subsubsection{Inclined Disks: No Gap}
\label{sec:inclnogap}

\begin{figure*}[htbp]
\plotone{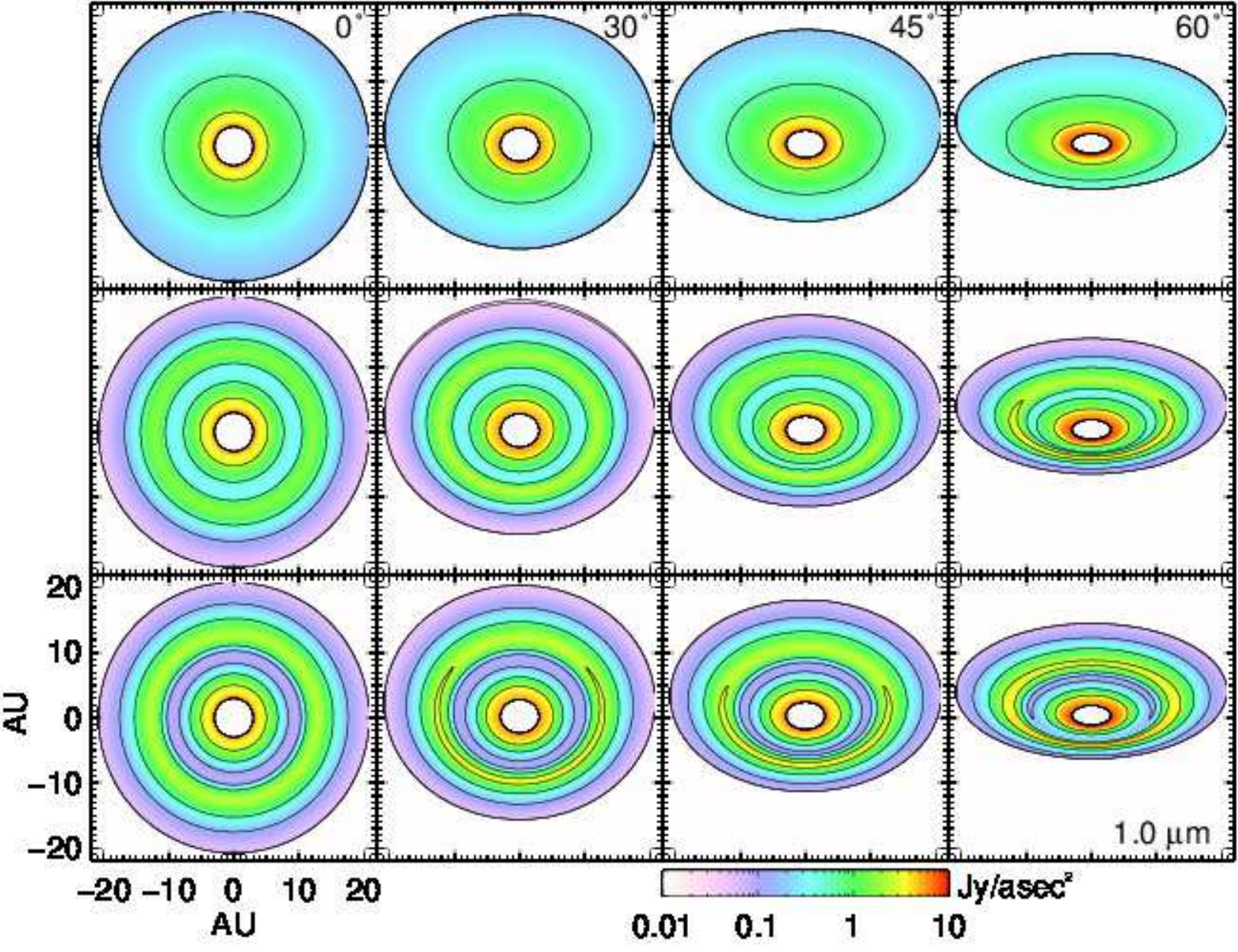}
\caption{\label{tiltedgaps0001um}
Images at 1 $\mu$m of disks with and without gaps viewed at 
various inclination angles.  
The top row of images show a disk without a gap.  
The middle (bottom) row shows images of a disk with a partial gap 
created by a 70 (200) $M_{\oplus}$ planet.  
From left to right, the inclination angles are 
$0\degr$, $30\degr$, $45\degr$, and $60\degr$.  
In each image, the upper part of the disk is tilted 
away from the observer.  
The scale for disk brightness is shown in the colorbar at the 
bottom.  
Contours are spaced by a factor of 4 in brightness. 
}
\end{figure*}

\begin{figure*}[htbp]
\plotone{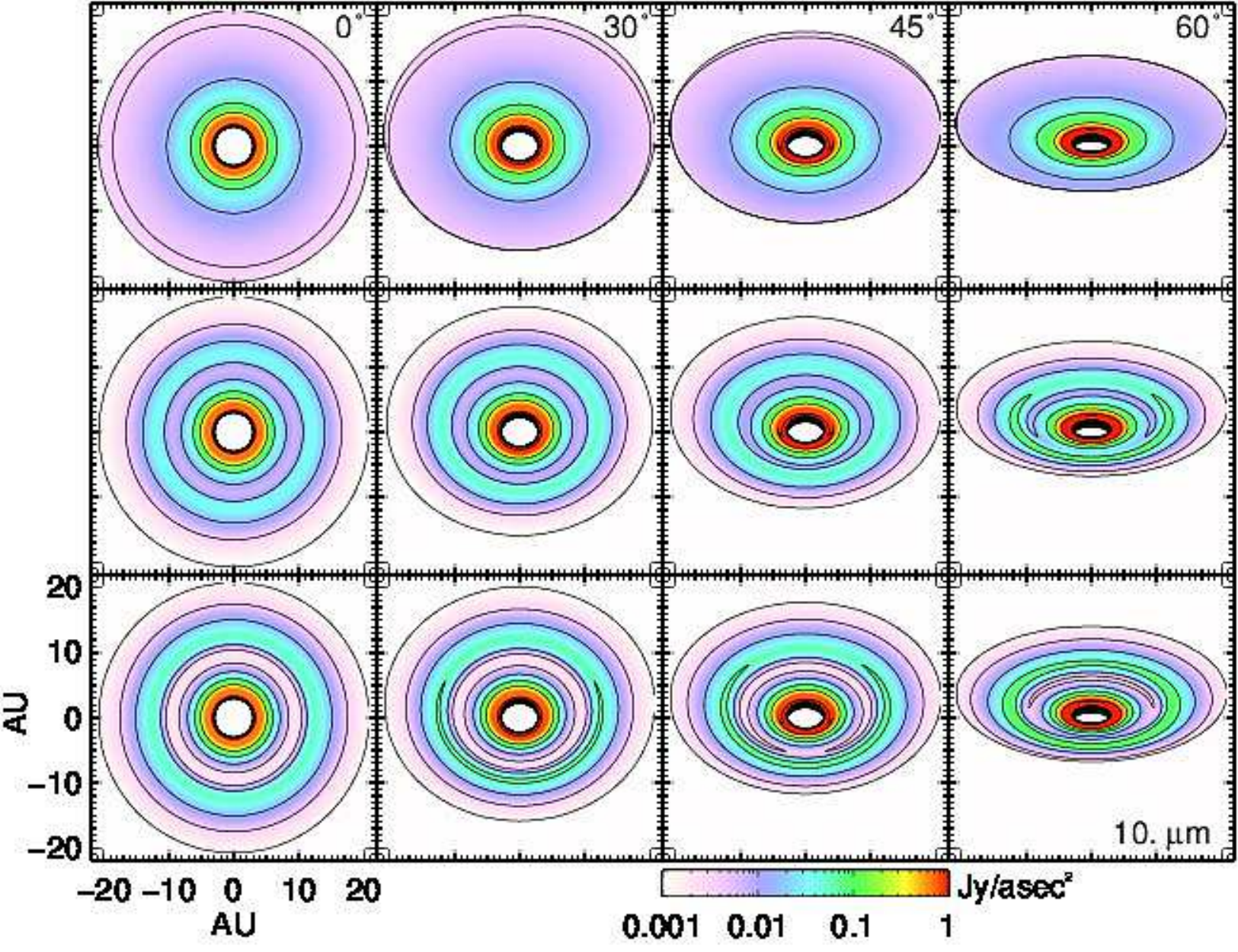}
\caption{\label{tiltedgaps0010um}
Same as Figure \ref{tiltedgaps0001um}, at 10 $\mu$m.  
Contours are spaced by a factor of 4 in brightness. 
}
\end{figure*}

\begin{figure*}[htbp]
\plotone{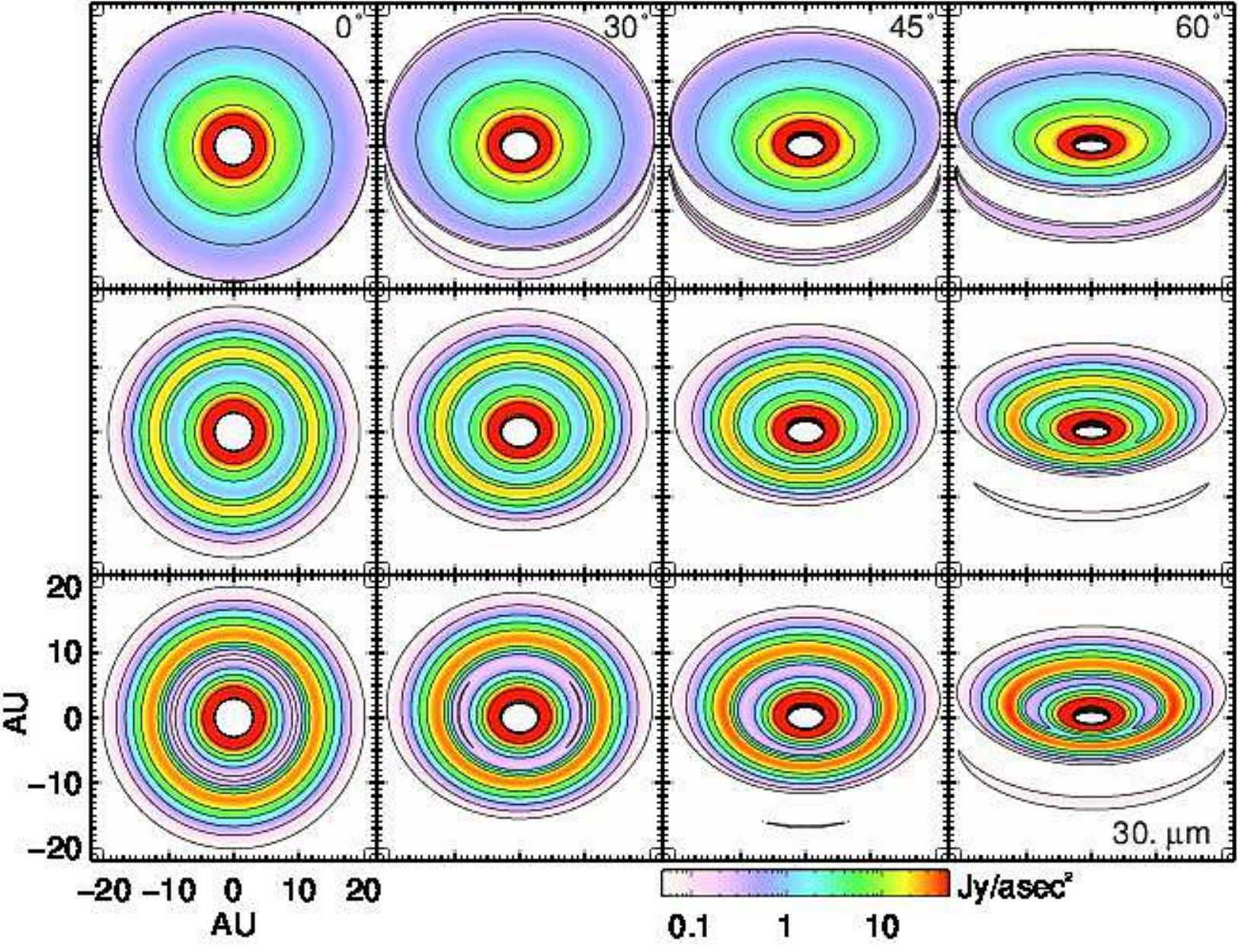}
\caption{\label{tiltedgaps0030um}
Same as Figure \ref{tiltedgaps0001um}, at 30 $\mu$m.  
Contours are spaced by a factor of 4 in brightness. 
}
\end{figure*}

\begin{figure*}[htbp]
\plotone{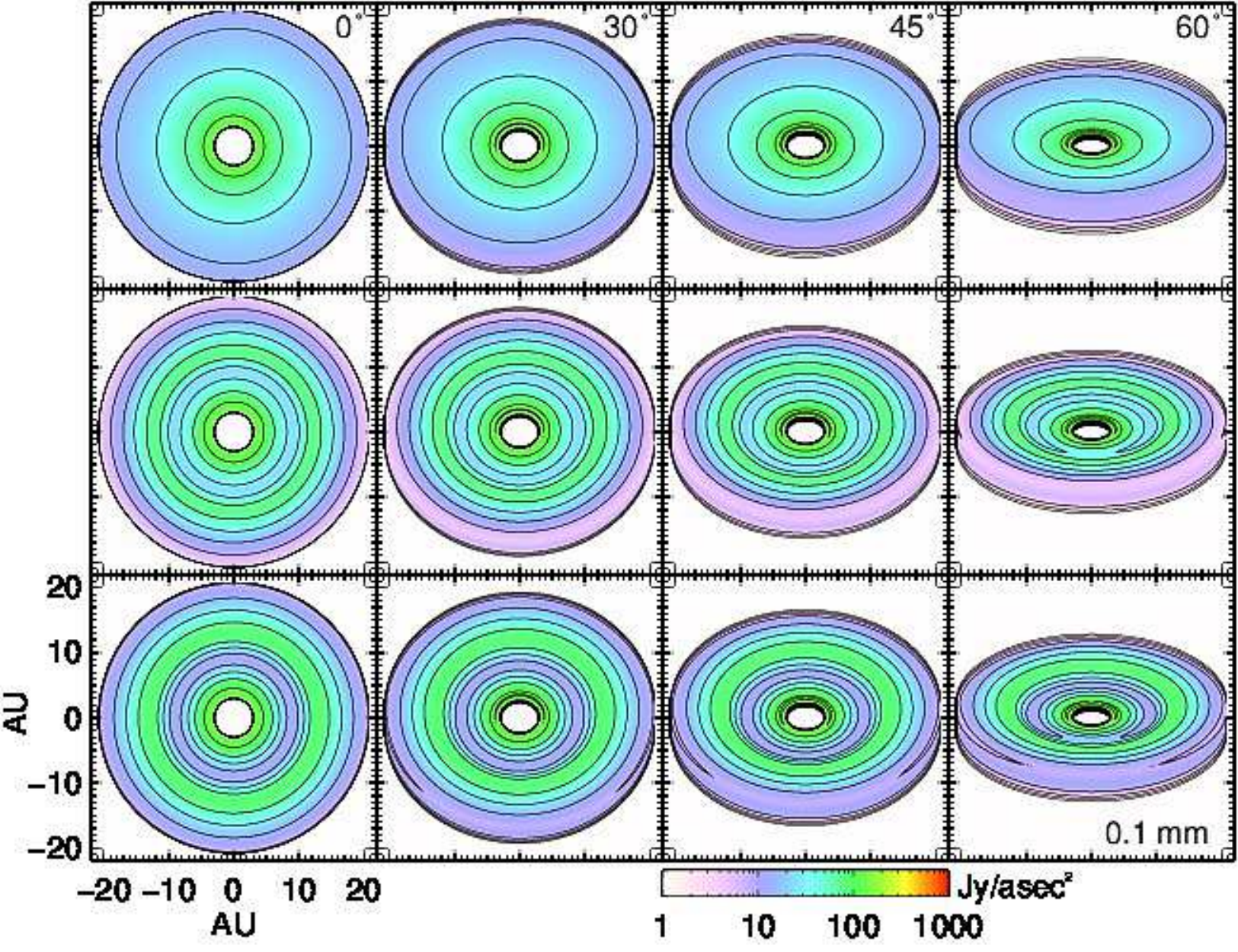}
\caption{\label{tiltedgaps0100um}
Same as Figure \ref{tiltedgaps0001um}, at 0.1 mm.  
Contours are spaced by a factor of 2 in brightness. 
}
\end{figure*}

\begin{figure*}[htbp]
\plotone{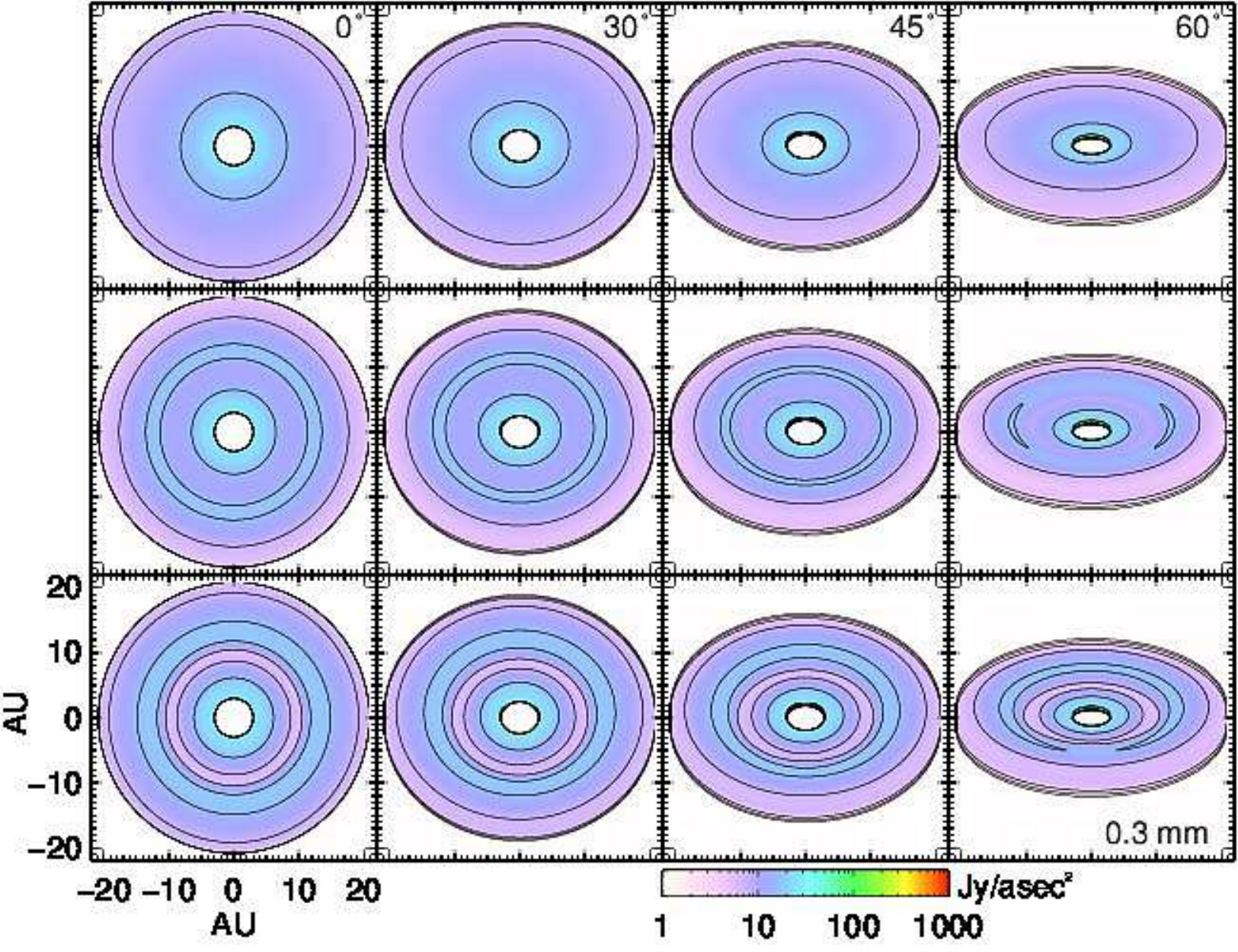}
\caption{\label{tiltedgaps0300um}
Same as Figure \ref{tiltedgaps0001um}, at 0.3 mm.  
Contours are spaced by a factor of 2 in brightness. 
}
\end{figure*}

\begin{figure*}[htbp]
\plotone{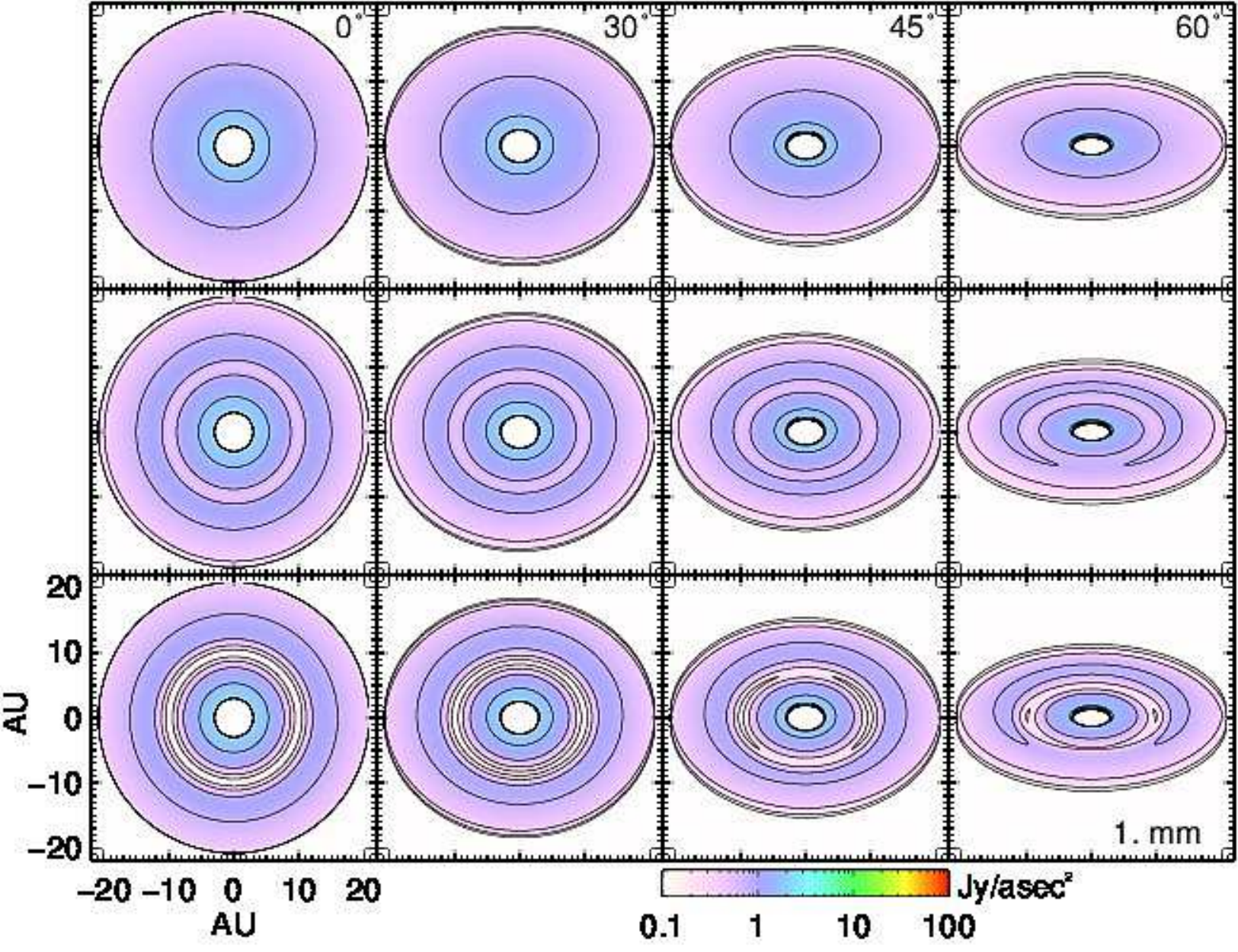}
\caption{\label{tiltedgaps1000um}
Same as Figure \ref{tiltedgaps0001um}, at 1 mm.  
Contours are spaced by a factor of 2 in brightness. 
}
\end{figure*}

In Figures \ref{tiltedgaps0001um}$-$\ref{tiltedgaps1000um}, 
we demonstrate the effect of inclination on the appearance disks 
when imaged at 1 $\mu$m to 1 mm, with and without gaps.  
The top row of each figure shows variously inclined disks 
without a gap.  
In each image, the disk is inclined so that 
the top edge tilted away from the observer.  
First, let us consider the fiducial disk model, in 
the absence of any gap.  

In the images at shorter wavelengths, the inclined disks 
appear vertically asymmetric because the emission 
is from the disk surface, which is essentially bowl-shaped.  
A schematic for the geometry is illustrated in \figref{tilteddisk}, 
which shows a cartoon schematic of a disk around a star.  The lines 
of sight marked by the angles 
$\eta_1$ and $\eta_2$ intersect the disk at the same 
distance from the star, but because the surface is above the midplane,
the projected distance on the sky is not equal, as can be 
seen by comparing the distance between the solid lines 
to the dashed line intersected in the star.  
This causes an apparent foreshortening of the near side of the disk.
Moreover, since 
$\eta_2>\eta_1$, the near side appears brighter than the far side.  
This is can be seen in the 1 $\mu$m images in the top row of 
\figref{tiltedgaps0001um}.  The foreshortening of the near side of the 
disk is also evident in the 10 and 30 $\mu$m images, 
in Figures \ref{tiltedgaps0010um} and \ref{tiltedgaps0030um}.  
At 0.1 mm and longer 
(Figures \ref{tiltedgaps0100um}, \ref{tiltedgaps0300um}, and 
\ref{tiltedgaps1000um}),
the optical depth decreases with increasing 
wavelength, so the vertical asymmetry diminishes.  The apparent 
thick rim on the disk's near side results
from truncating the calculation at 21 AU and should be ignored.

\begin{figure*}[htbp]
\centerline{\includegraphics{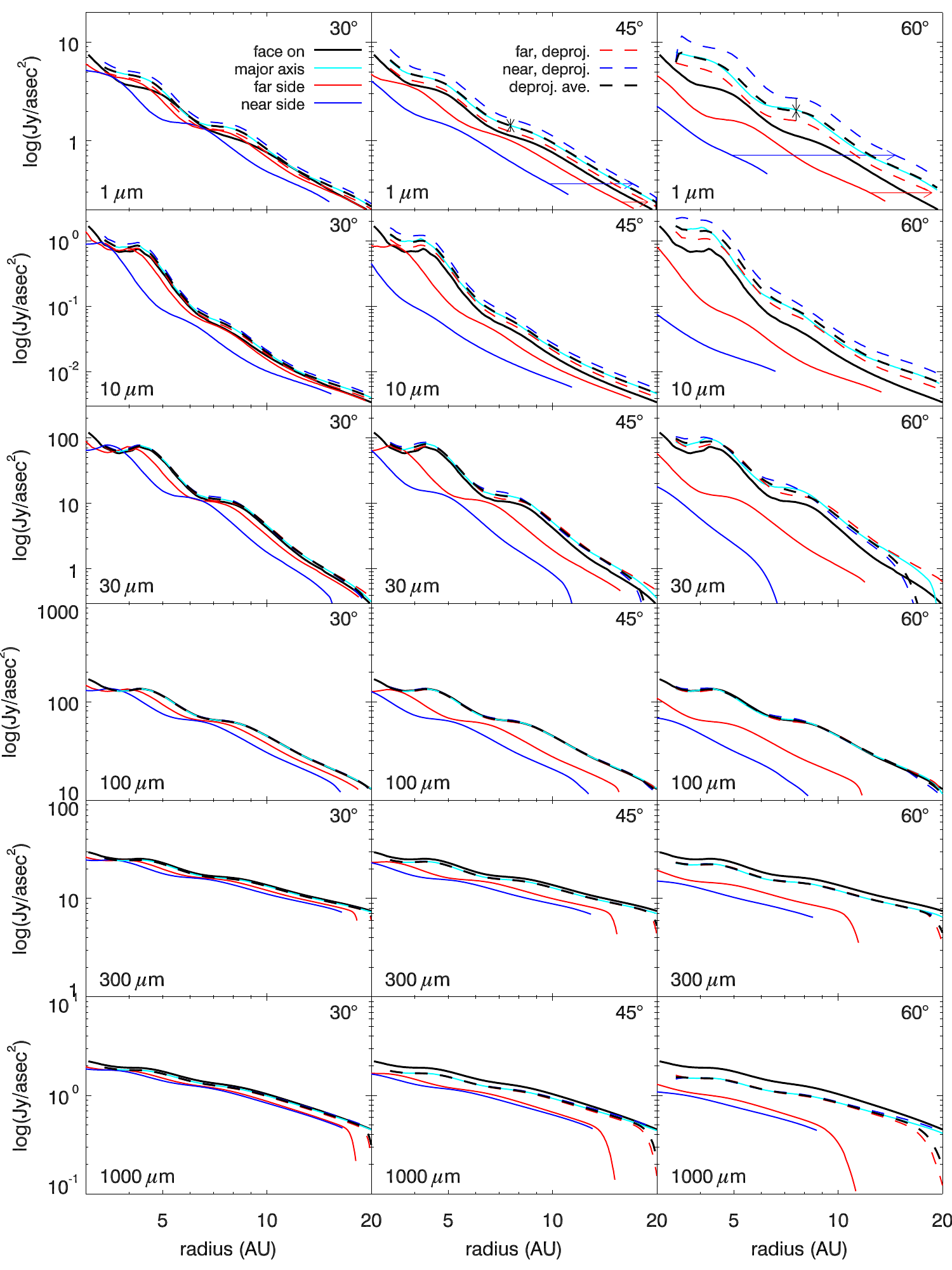}}
\caption{\label{nogapprofiles}
Radial surface brightness profiles of a disk without any gap
at 1, 10, 30, 100, 300, and 1000 microns, as indicated.  
The solid black line is the profile for a face on disk.  
The cyan, red, and blue lines show the profiles for a disk 
tilted at 45$\degr$, along the major axis, far minor axis, and near 
minor axis, respectively.  When the profiles along the 
minor axes are deprojected, accounting for both the inclination angle 
and the wavelength-dependent aspect angle, they shift to the 
red and blue dashed lines, respectively.  The geometrical average 
of these deprojected curves is plotted as the black dashed line.  
At all wavelengths, this average deprojected profile is a good 
fit to the major axis.  This relation may be an effective way to 
determine inclination angles of imaged protoplanetary disks.  
}
\end{figure*}

The surface brightness profile at 1 $\mu$m along the major 
axis of the tilted disk is systematically brighter 
as compared to the face on disk, as seen in the 
upper left plot of \figref{nogapprofiles}.  
The amount of brightening increases with increasing inclination angle.  
However,
this is not reproduced at wavelengths longward of 30 $\mu$m.  This can 
be explained as follows.  At 1 $\mu$m, the disk image is purely 
scattered light.  Thus, Equation (\ref{eq:totscat})
governs the brightness profile.  Assuming $\mu\ll1$, 
which is generally the case, then the brightness depends on the 
viewing angle $\eta$ as 
$\sim1/\cos\eta$.  
While Eq.~(\ref{eq:totscat}) is derived in detail in 
Paper I, another representation for the scattered light brightness 
is illustrated in \figref{schematic}.  
In essence, the star illuminates an optically thin layer 
of scatterers on the surface of the disk.  For $\mu\ll1$, 
the brightness is roughly proportional to the 
number of scatterers along the line of sight, which is in turn 
proportional to $1/\cos\eta$.  
When the disk is inclined at angle $i$, 
$\cos\eta = \cos\alpha\cos i$, along the 
major axis (at maximum elongation), as 
shown in Appendix \ref{app:angles}.  Since $\alpha$ is 
the angle of the disk 
surface with respect to the disk midplane and is intrinsic to the 
disk itself, the 
brightness profile along the major axis of the inclined disk 
scales as $1/\cos i$ 
as compared to the face on disk.  As $i$ increases, 
$1/\cos i$ increases as well.  

A similar effect occurs with the thermal emission to a limited extent.  
As seen in \figref{tempslices}, the surface of the disk is generally 
hotter than the optically thick interior.  At wavelengths 
close to the blackbody peak of the surface temperature, an inclination 
effect similar to that of the scattered light will occur.  
Since temperature decreases with distance in the disk at both the 
surface and the midplane, the wavelength of observation may be tuned 
to probe the surface at a particular distance in the disk.  
At 10 AU in this disk model, the surface temperatures are above 100K
while the interior is below 100K, so 30 $\mu$m imaging is a 
good probe of the surface of the disk. 
Wavelengths longer than 30 microns are sensitive to the interior 
of the disk both because the disk becomes optically thin 
and because the interior is cooler than the surface.  
This is illustrated in the disk profiles in \figref{nogapprofiles}, 
where the cyan lines indicating the major axis of the tilted disks 
are all systematically brighter than the face-on disk 
for $\lesssim 30$ microns,  
whereas at longer wavelengths the major axis profiles are 
less than or equal 
to the face-on profiles.  

In Table \ref{table:nogapincl} we list the heights and 
slopes of the disk surface at 10 AU at the listed wavelength.  
For wavelengths dominated by scattered light, we define 
the imaging surface, $z_s(r)$,  
to be where the disk becomes optically thick 
to stellar light at the given wavelength.
That is, $\tau_{\lambda}=2/3$ as measured along the line 
of sight to the star.  
At scattered light wavelengths (1 and 10 microns), 
$\mu \approx \partial z_s/\partial r - z_s/r$, 
and typically $\mu\ll1$.  
For thermal wavelengths ($\lambda\geq30\micron$), 
the surface is defined to be where 
the optical depth to observer in a face-on orientation 
becomes optically thick.  That is, 
$\tau_{\lambda}=2/3$ measured perpendicular to the disk.  

\begin{deluxetable*}{rcrrrrrrrr}
\tablecaption{\label{table:nogapincl}
Disk inclinations calculated from simulated images.}
\tablehead{
\colhead{$\lambda$} &
\colhead{emission} &
\multicolumn{2}{c}{actual\tablenotemark{$\dagger$}} &
\multicolumn{2}{c}{$i=30\degr$} &
\multicolumn{2}{c}{$i=45\degr$} &
\multicolumn{2}{c}{$i=60\degr$} 
\\
\colhead{($\mu$m)} &
\colhead{type} &
\colhead{$\partial z_s/\partial r$} & \colhead{$z_s/r$} &
\colhead{$i$} & \colhead{$z_s/r$} &
\colhead{$i$} & \colhead{$z_s/r$} &
\colhead{$i$} & \colhead{$z_s/r$} 
}
\tablecolumns{10}
\startdata
           1 & scattered
 &  0.221 &  0.185
 & $  29.8\degr$ & $ 0.185$
 & $  44.6\degr$ & $ 0.183$
 & $  59.3\degr$ & $ 0.191$
\\
          10 & scattered
 &  0.200 &  0.170
 & $  29.7\degr$ & $ 0.168$
 & $  44.5\degr$ & $ 0.171$
 & $  59.1\degr$ & $ 0.207$
\\
          30 & thermal
 &  0.112 &  0.109
 & $  29.9\degr$ & $ 0.172$
 & $  44.9\degr$ & $ 0.177$
 & $  59.7\degr$ & $ 0.177$
\\
         100 & thermal
 &  0.099 &  0.100
 & $  30.0\degr$ & $ 0.115$
 & $  45.2\degr$ & $ 0.104$
 & $  60.1\degr$ & $ 0.110$
\\
         300 & thermal
 &  0.081 &  0.085
 & $  30.2\degr$ & $ 0.047$
 & $  45.4\degr$ & $ 0.041$
 & $  60.5\degr$ & $ 0.050$
\\
        1000 & thermal
 &  0.052 &  0.066
 & $  30.5\degr$ & $ 0.002$
 & $  45.9\degr$ & $ 0.000$
 & $  61.0\degr$ & $ 0.003$
\enddata
\tablenotetext{$\dagger$}{Measured at $r=10$ AU}
\end{deluxetable*}

In addition to the brightness asymmetry, the projected distance 
at a given radial distance on the disk's surface is 
foreshortened on the near side of the disk compared
to the far side.  If $r_n$ and $r_f$ are the projected distances 
along the near and far sides of the disk, respectively, 
$r_d$ is the actual distance, 
and $\tan\beta(r) = z_s(r)/r$, then 
\begin{eqnarray}
r_n &=& r_d\cos(i+\beta)/\cos\beta \quad \mbox{and} \label{eq:rnear}\\
r_f &=& r_d\cos(i-\beta)/\cos\beta. \label{eq:rfar}
\end{eqnarray}
Note that these relations rely on the assumption that the 
imaging surface of the disk is independent of viewing 
angle.  This assumption is valid for scattered light, 
where the illuminated surface layer is thin compared to the depth to 
which the observer sees: that is, $\tau\sim1$ to the star occurs 
higher above the midplane than $\tau\sim1$ to the observer.  
This implies that the observer sees the entirety of the illuminated layer, 
although its brightness changes with viewing angle.  

Using Equations (\ref{eq:rnear})-(\ref{eq:rfar}) 
to rescale the radial distance, 
we can deproject the surface brightness profiles to 
bring them in line with what would be observed were the disk 
face-on.   
In \figref{nogapprofiles}, we plot the deprojected 
surface brightness profiles along the far and near 
sides of the disk as dashed red and blue lines, 
respectively.  In the 1 $\mu$m plot, the 
near side brightness profile appears brighter than the 
major axis, while the far side appears dimmer, as expected.  

The brightness asymmetry between the near and far sides of the disk, 
represented by red and blue lines in \figref{nogapprofiles}, 
can be quantified in terms of the disk geometry as well.  
Suppose the slope of the surface 
$\partial z_s/\partial r = \tan\alpha$ at $r=r_d$ and the disk is 
inclined at an angle $i$.  Let $\eta_1$ and $\eta_2$ be the 
the angles between the surface normal and the observer 
on the far side and near side, respectively.  
This is illustrated in \figref{tilteddisk}.  
By geometry, $\eta_1 = i-\alpha$ and $\eta_2 = i+\alpha$.  
A more general formulation for $\eta$ is discussed in 
Appendix \ref{app:angles}.  

Let $F_0(r_d)$ be the surface brightness of the face-on disk at 
the projected distance $r_d$.  For a tilted disk, the same 
three-dimensional radial distance is seen in projection 
at $r_d$, $r_n$, and $r_f$ 
along the major axis, near side minor axis, and far side minor 
axis, respectively.  The corresponding surface brightnesses are 
$F_m(r_d)$, $F_n(r_n)$, and $F_f(r_f)$.  
While these points are at the same physical distance from the 
star on the disk, they appear to be at different distances 
when seen in projection and appear to be different brightnesses 
because the values of $\eta$ are different in Eq.~(\ref{eq:totscat}).  
In particular, $F \propto 1/(\mu+\cos\eta)$.  
For the face-on disk, $\eta=\alpha$.  
Thus, we find that the brightness profiles along the 
major axis, near minor axis, and far minor axis are, respectively, 
\begin{eqnarray}
F_m(r_d)/F_0(r_d)&=&(\mu+\cos\alpha)/(\mu+\cos\alpha\cos i)
\label{scalemaj}\\
F_n(r_n)/F_0(r_d)&=&(\mu+\cos\alpha)/[\mu+\cos(i+\alpha)]
\label{scalenear}\\
F_f(r_f)/F_0(r_d)&=&(\mu+\cos\alpha)/[\mu+\cos(i-\alpha)].
\label{scalefar}\end{eqnarray}
Note that these relations rely on the assumption that 
the imaging surface of the disk ($z_s$) is locally plane-parallel, since 
the calculations for the brightnesses as detailed in section 
\S\ref{sec:observables} also rely on that assumption.  

Eq.~(\ref{scalemaj}) is the
harmonic mean of Eqs.~(\ref{scalenear}) and (\ref{scalefar}), so 
we find the relation
\begin{equation}\label{eq:harmmean}
\frac{1}{2} 
\left[\frac{1}{F_n(r_n)}
  +\frac{1}{F_f(r_f)}\right] = 
\frac{1}{F_m(r_d)}.  
\end{equation}
This means that given the correct values of $i$ and $\beta$ 
to deproject $r_n$ and $r_f$ according to Eqs.~(\ref{eq:rnear})-(\ref{eq:rfar}), 
then the harmonic mean of the deprojected 
surface brightness profiles on the near and far sides of the disk 
should equal the brightness profile along the 
major axis.  This harmonic mean is plotted as black dashed lines 
in \figref{nogapprofiles}, which does indeed prove
to be a good fit to the cyan line at all wavelengths.  

We can make use of this relation to construct a simple 
geometric model for a tilted disk image, and then 
apply this model to derive the inclination angle and thickness 
of a disk image.  
In order to reduce the number of free parameters, 
we make the assumption that 
$\partial z_s/\partial r \approx z_s/r$ or $\alpha\approx\beta$
and $\mu\approx0$.  
We determine the best fits for $i$ and $\beta$ over 
4 AU$<r_d<$20 AU by minimizing $\chi^2$ between the 
brightness profile along the major axis and the 
harmonic mean of the brightness profiles along the 
near and far minor axes, using the MPFIT package 
\citep{MPFIT}.\footnote{http://purl.com/net/mpfit}
In this way, we are effectively measuring 
$i$ and $\beta$ from anisotropies in disk images.  

The results of this fitting procedure are tabulated 
in the last two columns of Table \ref{table:nogapincl}. 
The fitting was done with no presumption about which side of the disk 
was near or far, but the correct orientation was still found.  
At all wavelengths and inclinations, the measured 
inclination angles are 
within 1\degr of the actual inclination.  
The measured values of $z_s/r$ 
are reasonably accurate for scattered light, but 
less so for thermal wavelengths.  
This is in part because Equations (\ref{eq:rnear})--(\ref{eq:harmmean})
were derived from the equations for scattered light images.  
For example, the assumption that the imaging surface is independent of 
viewing angle is a reasonably good approximation for scattered light, 
since the illuminated layer of the disk is independent of viewing 
angle.  However, at thermal wavelengths, the height of the imaging 
surface does depend on viewing angle, so that unless the transition 
from optically thin to thick is narrow, Equations 
(\ref{eq:rnear})--(\ref{eq:rfar}) do not obtain.  
Nevertheless, we find that we can successfully 
recover the inclination angle of the disk, even if we cannot 
measure the aspect ratio of the disk ($z_s/r$) well.  

Using Equations (\ref{eq:rnear})--(\ref{eq:harmmean}) 
to fit for $i$ and $z_s/r$ only works well if 
$z_s/r$ varies slowly with $r$.  Otherwise fitting to a 
a single value of $\alpha$ is inaccurate.  
Another effect that was not considered for these models was 
forward-scattering of light by small dust grains.  
A prediction of Mie theory is that small dust grains 
can be strongly forward-scattering, but scattering is assumed 
to be completely isotropic in the models presented here.
On the other hand, if the anisotropy of the scattering 
can be well-described by a single parameter, such as in the 
Henyey-Greenstein model \citep{HenyeyGreenstein}, 
the anisotropy may be treated as an additional fitting parameter 
and solved for accordingly.  However, this is outside the scope 
of this paper.  

\subsubsection{Inclined Disks: With Gaps}
\label{sec:gapimages}

Simulated images of inclined disks with gaps imposed on them by 70 
and 200 $M_{\oplus}$ planets at 10 AU from the star 
are shown in the bottom two rows of 
Figures \ref{tiltedgaps0001um}$-$\ref{tiltedgaps1000um}.  
As also seen in \citet{2006Varniere_etal}, the far shoulder 
of the gap is brightened in scattered light (Figure \ref{tiltedgaps0001um}) 
regardless of inclination angle.  

In the images of tilted gapped disks, the asymmetry in brightness between 
the near and far sides of the disk is more apparent than in the 
gapless disk.  From 1 to 30 microns, 
the brightened shoulder of gap is brighter on the lower 
half of the disk, the side that is tipped toward the observer.  
In addition, the width of the shadow within the gap is narrower 
on the near side because of the geometric foreshortening of the disk.  
At 100 microns and beyond, these asymmetries become less apparent 
as the disk becomes optically thin.  

We can again make use of Eqs.~(\ref{scalemaj}-\ref{scalefar}) 
to estimate the brightness of the shadow in the gap and the brightened 
far shoulder as a function of PA.  
The gap in the disk makes the determination of disk thickness and 
inclination angle easier, because the shadow and brightening in the gap 
are points of reference that we can use to compare the profiles along 
different PA to each other.  

\begin{figure}[htbp]
\plotone{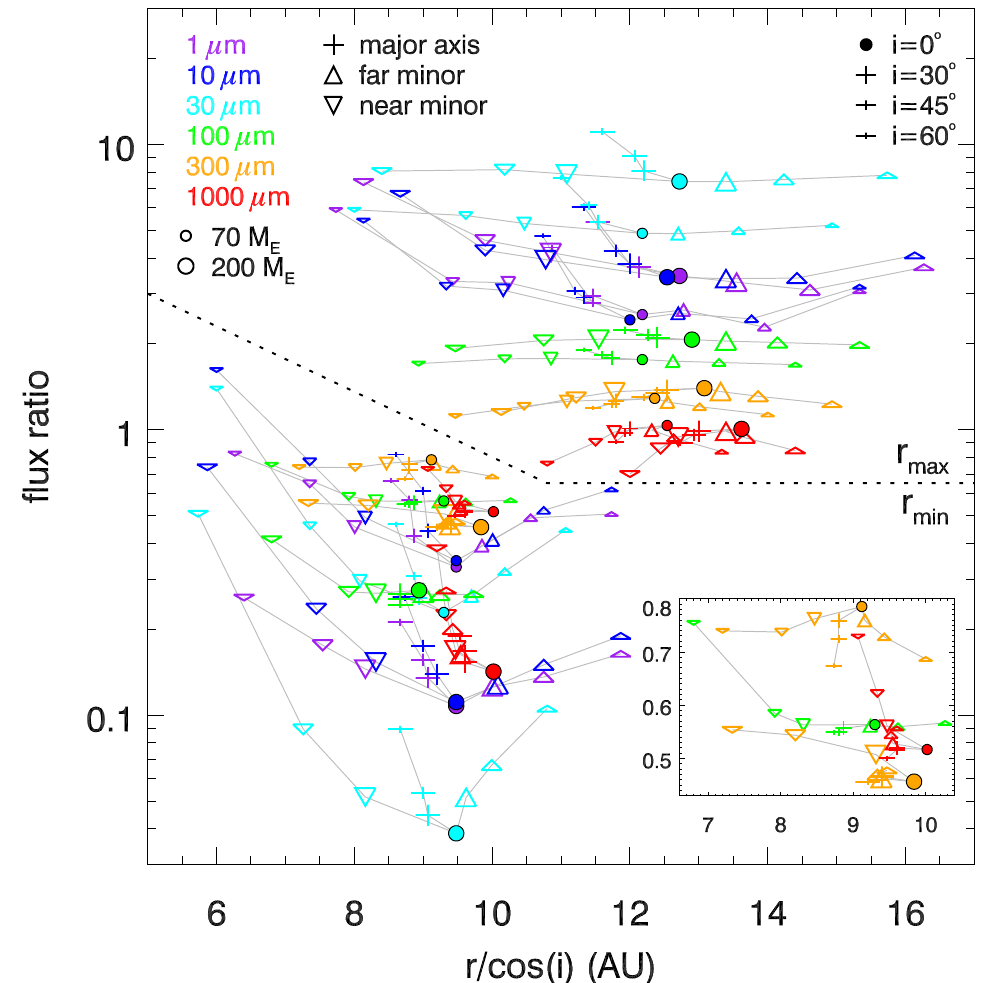}
\caption{\label{gapmaxmin}
Flux ratios versus positions of brightness maxima and minima 
for disks with gaps at 10 AU, observed at varying inclination angle.  
The flux ratio is the brightness 
maximum or minimum divided by the brightness of 
a gapless disk at a radius 
of 10 AU from the star when observed at $i=0\degr$.  
Brightness maxima/minima are plotted above/below the dotted line, and 
are measured along a cut through the 
disk image going through the star and parallel to either the major ($+$) or 
minor ($\triangle$/$\bigtriangledown$ for far/near side)
axis.  Circles indicate disks with $i=0\degr$, 
and the aspect ratio of the symbol decreases with increasing 
inclination angle.   Gray lines connect model images 
varying with inclination angle.  
Along the minor axis, the radius is scaled by $1/\cos i$.  
The wavelength of observation is indicated by 
color: purple (1\micron), blue (10\micron), 
cyan (30\micron), green (0.1 mm), orange (0.3 mm), and red (1 mm).  
Smaller/larger points are used for gaps opened by 70/200 $M_{\earth}$
planets.  The inset clarifies the region of 
6.5 AU $<r\sub{min}<$ 10.5 AU.  
}
\end{figure}

Along the major axis, near minor axis, and far minor axis, we 
measure the surface brightness profile at each wavelength of observation 
as a function of projected 
radius.  For each of these surface brightness profiles, we find the 
local minimum in brightness caused by the gap, and the local maximum on the 
far gap shoulder, labeling these points $r\sub{min}$ and $r\sub{max}$, 
respectively.  We plot the flux ratio at these points versus 
$r\sub{min}$ and $r\sub{max}$ in \figref{gapmaxmin}, where 
the flux ratio is defined to be 
the brightness at an extremum scaled to 
the disk brightness at 10 AU on the gap-less disk seen face-on.  
Variation of brightness extrema with wavelength, 
projection axis, planet mass, and inclination are indicated by the color, 
shape, size, and aspect ratio of symbols, respectively.  

Before discussing the derived values of $i$ and $\beta$ 
from measurements of $r\sub{max}$ and $r\sub{min}$, we shall 
examine the variation of $r\sub{max}$ and $r\sub{min}$ and 
the brightnesses at these points as a function of 
wavelength, planet mass, and inclination angle.  
Each circle in \figref{gapmaxmin} represents the radius 
and flux ratio of the brightness extrema 
in a face-on disk for each of the two modeled planet masses 
as observed at a specific wavelength.  The three branches 
coming off each circle indicate how the position and 
brightness of the extrema vary with along the major and minor axes 
as the inclination angles changes.  Along the minor axis, 
$r\sub{max}$ and $r\sub{min}$ are scaled by 
$1/\cos i$ to partially account for inclination angle.  

In nearly all cases, $r\sub{max}/\cos i$ and $r\sub{min}/\cos i$ 
is smaller on the near side than on the far side, with the values 
along the major axis somewhere in between.    
This is illustrated by the far side 
branches (marked by triangles, $\triangle$) lying to the right of the 
major axis branches ($+$), and near side branches ($\bigtriangledown$) 
lying to the left in \figref{gapmaxmin}.  This is because the 
imaging surface is well above the midplane, creating additional 
foreshortening of the disk on the near side, and less on the far side 
of the disk.  The few exceptions to these general trends occur 
at the longest wavelengths, 0.3 and 1 mm.  

Additionally, $r\sub{max}$ and $r\sub{min}$ is smaller along the 
major axis of an inclined disk as compared to the $i=0$ image 
in almost all cases.  This is also due to the thickness of the disk.  
If one were to consider the ellipse traced by brightness maxima and 
minima along all azimuthal angles, the center of this ellipse 
would be offset from the position of the star, toward the direction 
of the far side.  Because of this, a cut through the image through 
the star parallel to the major axis would be offset from the major 
axis of the ellipse, thus $r\sub{max}$ and $r\sub{min}$ as measured 
for \figref{gapmaxmin} are smaller along the major axis as 
compared to a face-on image. 

The flux ratios of brightness minima have a larger spread
than the brightness maxima, particularly for $\lambda\leq0.1$ mm. 
This is because 
tilting of the disk obstructs the shadowed region, decreasing 
the amount of contrast, particularly on the near side of the disk.  
At longer wavelengths, the disk becomes optically thin, and geometric
brightening is no longer important.  

The brightnesses along the major axis for $\lambda \lesssim100$ $\mu$m 
are generally higher than for the face-on disk, increasing with $i$, because 
$\cos\eta = \cos\alpha\cos i$ (see Appendix \ref{app:angles}).  
That is, in scattered light, the viewing angle 
increases with increasing inclination angle, causing 
the disk to appear brighter.   
At longer wavelengths, on the other hand, we expect some dilution of 
brightness extrema as our line of sight through the disk changes, 
though the effect is small because of the decreasing opacity of the 
disk toward long wavelength.  Hence, the brightness maxima become 
dimmer, and the brightness minima become brighter for 
$\lambda\gtrsim0.3$ mm.  
Although images at 30 and 100\micron 
are of thermal emission, that emission comes from closer to the 
surface of the disk than at 0.3 or 1 mm, so there is less dilution 
of brightness along the line of sight, and the behavior with inclination 
angle is shares similarities to that for scattered light.
The exception to these patterns are the brightness minima 
at $0.1-1$ mm (see inset of \figref{gapmaxmin}).  
In some cases, the brightness minima along the major 
are dimmer as inclination increases, most notably for the 
70 $M_{\earth}$ planet observed at 0.3 mm.  This is caused 
by the overall dimming of the disk with increasing inclination angle,  
also seen in \figref{nogapprofiles}.

We find that the near side of the disk is systematically 
brighter than the far side at $r\sub{min}$ and $r\sub{max}$ than along 
the major axis for $\lambda\leq30\micron$, as expected because 
$\eta$ is larger on the near side than on the far side.    

\begin{figure}[tbh]
\plotone{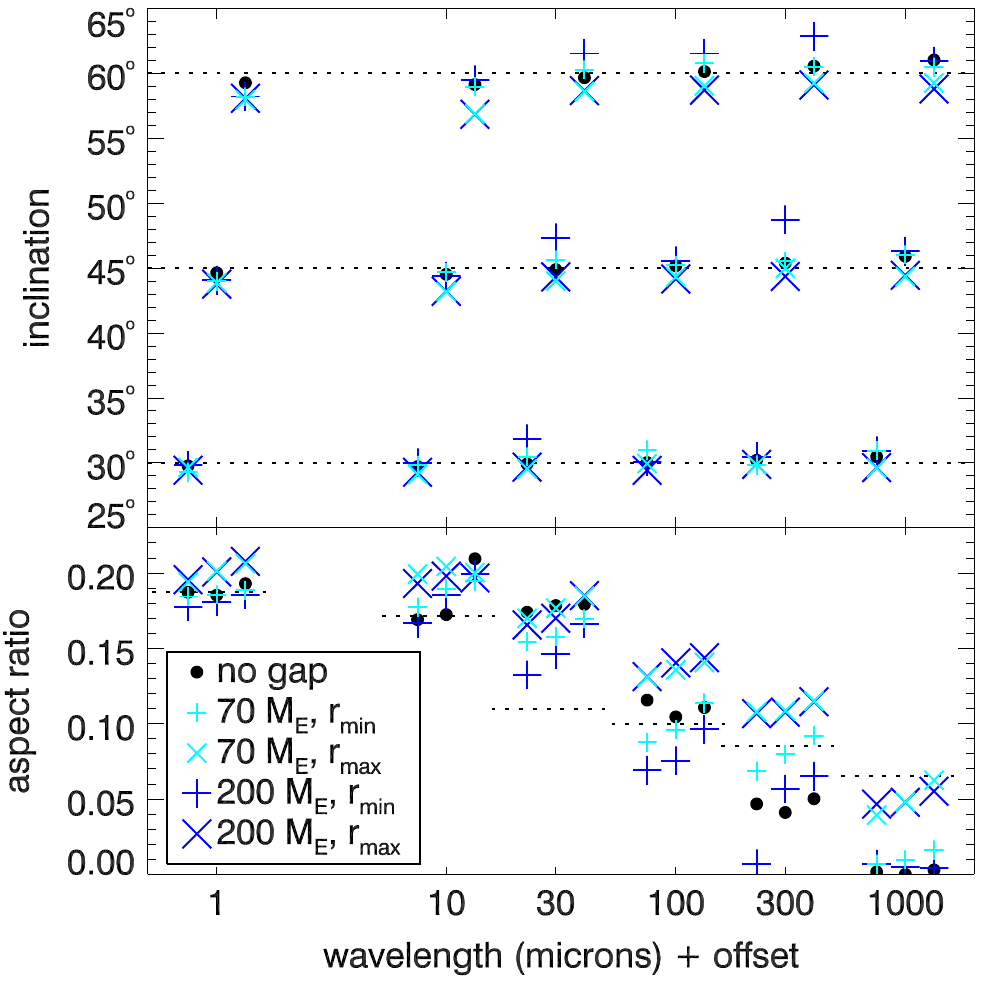}
\caption{\label{fitangles}
Inclination angles (top panel) and aspect ratios (bottom panel) 
calculated from simulated disk images 
as compared to actual values.  
The aspect ratio here is defined as $z_s/r = \tan\beta$.  
The horizontal axis is the wavelength of observation plus 
an offset corresponding to inclination angle, so that points from 
left to right have $i=30\degr,45\degr,$ and $60\degr$. 
Actual values are indicated by horizontal dotted lines.  
Black circles indicate values tabulated in 
Table \ref{table:nogapincl} for disks without gaps.  
Crosses ($\times$) and plus symbols ($+$) 
indicate values derived from $r\sub{max}$ and $r\sub{min}$ 
measurements, respectively.  
Smaller cyan symbols indicate 
a disk model with a 70 $M_{\earth}$ planet, while 
larger blue symbols indicate a model with 200 $M_{\earth}$.  
}
\end{figure}

As in \S\ref{sec:inclnogap}, we can use the geometric relations 
between $r_d$, $r_n$, and $r_f$ to back out the inclination and 
aspect angles of the disk.  
From Eqs.~(\ref{eq:rnear}-\ref{eq:rfar}), 
\begin{eqnarray}
\frac{1}{2}\left(r_f + r_n\right) &=& r_d \cos i \quad\mbox{and} \label{eq:gapincl}\\
\frac{1}{2}\left(r_f - r_n\right) &=& r_d \sin i \tan\beta 
. \label{eq:gapaspect}
\end{eqnarray}
Using the measured values of $r\sub{min}$ and $r\sub{max}$ 
along the near and far minor axes as 
references point, we can then solve for the inclination ($i$) 
and aspect ratio ($z_s/r=\tan\beta$)
of the disk from Eqs.~(\ref{eq:gapincl}-\ref{eq:gapaspect}).  
These derived values from simulated disk images are plotted in 
\figref{fitangles}. 

For reference, values of $i$ and $z_s/r$ as calculated from 
full radial profiles of the fiducial disk model 
and tabulated in Table \ref{table:nogapincl} 
are also plotted in \figref{fitangles}.  
Comparing these values to those plotted in \figref{fitangles}, 
we find that although it may be 
easier to calculate $i$ and $z_s/r$ by using the 
brightness maxima and minima as reference points, 
the values returned are not necessarily more accurate than 
those obtained by comparing the overall surface brightness profiles.  
Better accuracy can be obtained by averaging the values of 
$i$ calculated from $r\sub{max}$ and $r\sub{min}$, although 
estimates of $i$ are less accurate for the larger planet mass.  

Estimates of the aspect ratio of the disk are reasonably 
accurate in scattered light (1 and 10 $\mu$m), 
as seen in the lower panel of \figref{fitangles}.  
However, at thermal wavelengths ($\lambda\geq30\micron$), 
measurements of $z_s/r$ from either radial profiles of 
gapless disks or brightness extrema in disks with gaps 
are less accurate in general.  Both the measured and true values 
of $z_s/r$ decrease with increasing wavelength, but the 
relation is steeper for the measured values.  
Thus, measurements at 30 $\mu$m all over-estimate $z_s/r$, 
while measurements at 1 mm all under-estimate $z_s/r$.  
Interestingly, the measured values are fairly 
consistent with each other across inclination angle, 
although there is a trend toward higher measured values of 
$z_s/r$ with increasing inclination.  
This suggests that using a $\tau=2/3$ height for the 
disk is not what is being probed by thermal emission.  
For example, at 30 $\mu$m, we are likely seeing emission from the 
hotter surface layer of the disk rather than to a $\tau=2/3$ depth.  
Since this layer is higher above the midplane than the 
$\tau=2/3$ depth, the measured aspect ratio is higher.  
At longer wavelengths, a likely explanation for 
the mismatch in aspect ratio is that thermal emission 
comes from broadly throughout the disk, as opposed to a 
vertically confined layer as in scattered light.  

Finally, we note that our results hold only for 
circular gaps centered on the star.  
If the gap is eccentric, as might 
be produced by an eccentric planet, additional errors in the fitting 
will be introduced, and it will be difficult to disentangle 
changes in morphology produced from eccentricity versus 
inclination angle and disk thickness.

\section{LkCa 15}

To demonstrate the application of the simulated gapped disk images to 
real observations, we 
examine the case of LkCa 15.  
LkCa 15 is a T Tauri star that has been identified as having a 
transitional disk because it appears to have an inner cavity 
of radius 46 AU 
as inferred from its SED \citep{2007Espaillat_etal}.  
The inner cavity has also been 
directly imaged by radio interferometry 
\citep{2011Andrews_etal,2011Andrews_LkCa15,2006Pietu_etal,2009IsellaCarpenterSargent}, 
and in scattered light 
\citep{2010Thalmann_etal}.  
LkCa 15 is also sometimes referred to as a ``pre-transitional'' disk 
because the inner cavity is not completely cleared 
\citep{2008Espaillat_etal}.  

One enticing possibility for the clearing of the 
inner disk in LkCa 15 is planet formation.  
\citet{2010Pott_etal} observed no stellar companion 
down to 3.5 AU separations from the star.   
Recently, a possible planet has been imaged in its central cavity, 
at a deprojected distance of $\sim20$ AU 
\citep{2012KrausIreland}, but it is too far from the 
wall at 46 AU to be solely responsible for the inner clearing.  
\citet{2010Bonavita_etal} put an upper limit of 
5 $M\sub{Jup}$ on a possible companion using 
NACO observations.  From a theoretical standpoint, 
planets more massive than 6 $M\sub{Jup}$ should cut off 
any accretion onto the star \citep{lsa99}, 
but \citet{2007Espaillat_etal} find that the disk 
is still accreting onto the star at a rate of 
$\dot{M}=2.4\times10^{-9}\,M_{\odot}\,\mbox{yr}^{-1}$.  
Gas has been detected inside the cavity in the form of 
CO lines \citep{2007Pietu_etal}, indicating an incompletely 
cleared inner disk.  
LkCa 15 also boasts a warm dust component at 0.12-0.15 AU
as inferred from its near-infrared excess 
\citep{2008Espaillat_etal}, further evidence that the inner 
cavity is not completely cleared.  
All these lines of evidence point to the suggestion that 
one or more planetary-mass companions are responsible for the 
inner cavity in LkCa 15, with an upper limit on a single 
planet of 5 $M\sub{Jup}$.  

Using the methods outlined above, we can compare simulated images of 
gaps in disks created by planets of varying mass to the scattered 
light image of LkCa 15.  
Supposing that a single embedded planet is responsible for the inner 
clearing in LkCa 15, we can then put a constraint on the mass 
of a possible planet in LkCa 15.  
Although the models presented in this paper obtain for 
a radially contrained gap in a disk rather than a full inner 
clearing of all the material interior to a given radius 
within a disk, comparing the masses of planets capable of creating 
only gaps or partial gaps gives a useful lower bound on the 
mass of a possible planetary companion to LkCa 15.  

LkCa 15 has a stellar mass of 
$M_*=0.97 \pm 0.03\,M_{\odot}$, 
effective temperature 
$T\sub{eff} = 4350$ K, 
and luminosity $L_*=0.74\,L_{\odot}$
\citep{2000Simon_etal}.
Its disk is inclined at approximately 52\degr \citep{2007Pietu_etal} 
and its 
properties are well-fit with 
an accretion rate of 
$\dot{M} = 2.4\times 10^{-9}\,M_{\odot}\mbox{yr}^{-1}$ and 
$\alpha\sub{ss} = 0.0007$ \citep{2007Espaillat_etal,2010Thalmann_etal}.  
For the purposes of this study, we round the stellar mass to 
$1\,M_{\odot}$ and derive a 
stellar radius of $R_* = 1.5\,R_{\odot}$ from 
the effective temperature and luminosity.  
These parameters are also listed in Table \ref{table:params}.
We calculate a new disk model using the same procedures 
as for the fiducial model described above, 
but changing the parameters for the star and disk as constrained 
by observations.  
To simplify the calculation, 
the initial surface density 
profile is simply interpolated from a locally-plane parallel 
disk model rather than recalculating the full structure in detail.  
The radial range of this slice is 
from 9.5 to 99.5 AU, and the vertical range is 0 to 28.9 AU.  

\begin{figure}[htbp]
\plotone{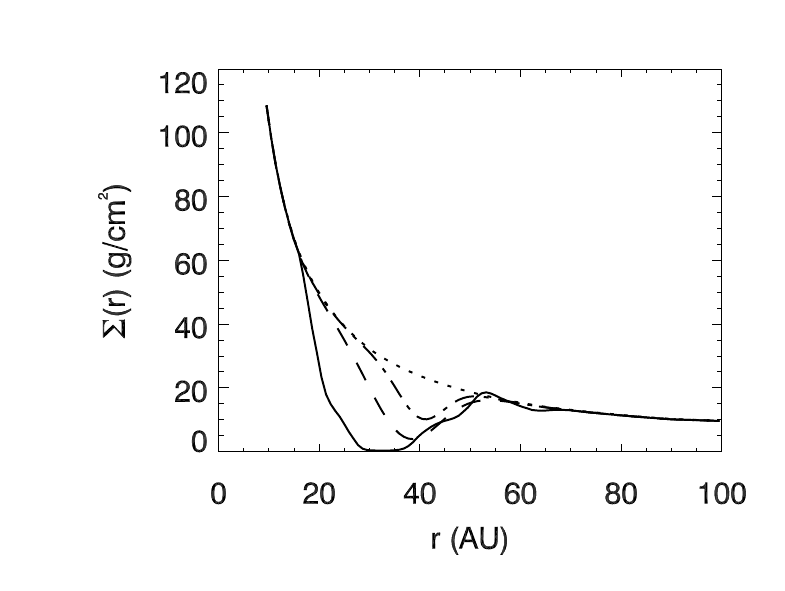}
\caption{\label{LkCa15density} 
Surface density profiles of the disks used to model LkCa 15. 
The dotted line shows the unperturbed profile with no planet.  
The dot-dashed, dashed, and solid lines show gaps created by 
planets of 11, 37, and 154 $M_{\oplus}$ at 
40.7, 38.3, and 32.5 AU, respectively.  
}
\end{figure}

Since the width of a gap created by a planet varies with planet mass, 
a larger planet would be located further from the gap wall than a 
smaller planet.  Thus, for a given gap size, we position it so that 
its half depth on the far side is at 46 AU.  That is, we 
position the gap so that 
\begin{equation}\label{gapeq}
\frac{\Sigma(r=46\mbox{ AU})}{\Sigma_0(r=46\mbox{ AU})} = 1/2
\end{equation}
(where $\Sigma$ without a subscript is the density 
profile with a gap and $\Sigma_0$ is the unperturbed density) 
with the gap trough interior to this distance. 
As in Paper I, we define the gap-opening threshold to be when 
$G=1$, where $G$ is the gap-opening parameter defined 
in Eq.~\ref{eq:viscgapcrit}.  
A planet with $G=1$ 
opens a gap that is not well-modeled with Equation (\ref{gapeq}).  
\citet{bate} calculate the surface density 
profile of a $1 M_J$ planet with $G=1.04$ in their hydrodynamic 
simulations, and we adopt this as an axisymmetric gap profile 
without fitting to a Gaussian.  
The calculated gap parameters are summarized in 
Table \ref{table:params}
and their surface density profiles 
are plotted in \figref{LkCa15density}.  

\begin{figure*}[htbp]
\plotone{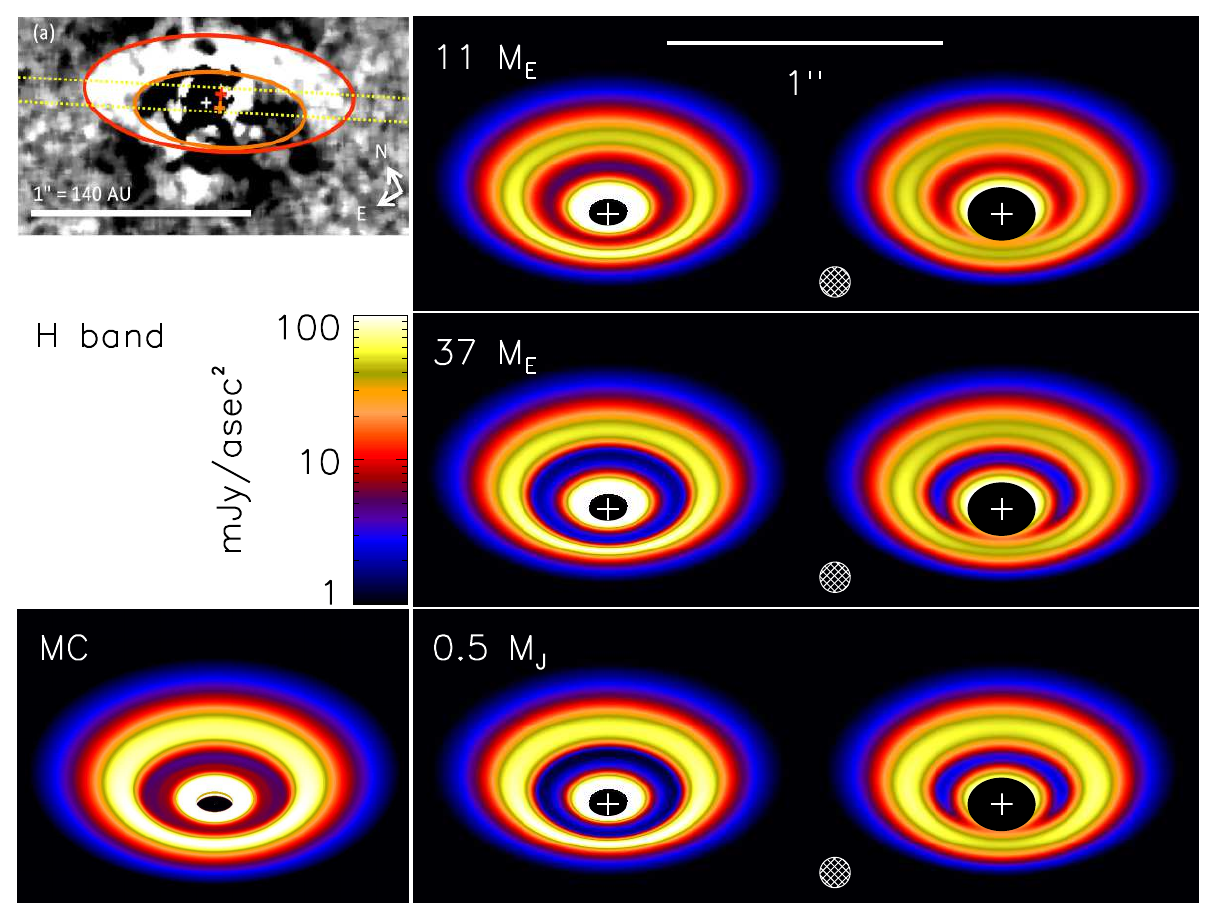}
\caption{\label{lkca15H}
Observed and synthetic H-band images of the disk around LkCa 15.  
Upper left: H band image by \citet{2010Thalmann_etal}, 
with the scale bar representing 140 AU.  
Remaining images are synthetic images based on models for LkCa 15.  
The disk is inclined at 52\degr and oriented so that the 
top edge of the disk is tipped away from the observer.  
The white cross indicates the position of the central star, 
and the blacked out inner region is outside the simulation 
boundaries.  
The gaps in the models are created by 
11 (top row), 37 (middle row), and 150 (bottom row) $M_{\earth}$ planets.  
The images in the center column are the idealized JC models, while 
the images on the right have been convolved 
with a Gaussian PSF of FWHM of 0\arcsec.055, as represented by 
the size of the hashed white circles.  
The lower left image is the MC image of the 150 $M_{\earth}$ 
model, generated using $10^8$ photon packets.  
}
\end{figure*}

In \figref{lkca15H}, we show observed and 
simulated H-band images of LkCa 15.  
The upper left panel is the H-band Subaru image 
obtained by \cite{2010Thalmann_etal}, and the 
remaining panels are simulated images of our LkCa15 disk 
model with an 11 $M_{\oplus}$, 37 $M_{\oplus}$, and 0.5 $M_J$ planet.  
The model disks are oriented 
so that the top part of the disk is the far edge, 
and the near edge is at the bottom.  
The center column of images are the fully resolved disk images, while 
the right column shows the images convolved with a Gaussian PSF
of 0.55\arcsec, consistent with the resolution of the 
Thalmann, et al.~image.  
The lower left image is the MC prediction for the 0.5 $M_J$ planet case, 
using $10^8$ photon packets.  The morphology of the MC image is 
nearly identical to the JC image at bottom center of \figref{lkca15H}.  
However, as discussed earlier, the gap shadow contrast is less in the 
MC image.  All models over-predict thermal 
emission from the inner disk in the SED, 
indicating that the extent of disk clearing is wider than 
a gap created by a single planet of 0.5 $M_J$.

The brightness anisotropy in the model images 
is solely due to geometric effects.  
The point of maximum brightness is on the near side of the disk, but
since the near side has a smaller angular size, when it is convolved
with a realistic telescope resolution, the far side appears brighter
overall.  Thus, the model prediction is that the bright 
disk emission seen in the H-band image obtained by 
\citet[][see top left panel of \figref{lkca15H}]{2010Thalmann_etal} 
is from back-scattering off the inner wall of the outer disk. 

The blacked-out inner circle in the simulated images in the right 
column of \figref{lkca15H} represents the 0\arcsec.055 FWHM PSF 
as reported in \citet{2010Thalmann_etal}.  
It is difficult to determine 
which model image (37 $M_{\oplus}$ or 0.5 $M_J$) is a better 
match to the observation, because they differ primarily in 
brightness, and the method used by 
Thalmann et al.~to gain high contrast in order to image the disk 
does not preserve total flux.  However, both models over-predict 
emission from the inner disk compared to the observational evidence.  
In our simulated images, the inner disk is quite bright and extends 
past the PSF circle, but this is not seen in the Thalmann
et al.~observation.  
Moreover, detailed analysis of LkCa 15's spectrum indicates that 
the dust in the inner disk extends no further than 5 AU 
\citep{2008Espaillat_etal}.  
This suggests that either the gap in LkCa 15 is caused by
a single planet more massive than 0.5 $M_J$, or there are 
multiple planets in the gap.

Using the red and orange ellipses drawn on the top left panel 
of \figref{lkca15H}, we can attempt to use the relations 
given in Eq.~(\ref{eq:gapincl})-(\ref{eq:gapaspect}), 
using the midpoints between 
the two ellipses.  Then, 
$r_f$, $r_n$ and $r_d$ are 28, 18, and 55 AU, respectively.  
This gives an inclination of 65\degr and $z_s/r=0.1.$  
For comparison, our model has an assumed inclination of 52\degr, 
and $z_s/r=0.2$ at 46 AU\@.  
However, the values measured from the observations 
are highly uncertain for two reasons: 
(1) the LOCI algorithm used for revealing the disk does not conserve flux 
so the measurements are highly uncertain themselves, and 
(2) the disk may be intrinsically elliptical, as evidenced by the 
offset of the star from the center of the major axis, in which 
case Eqs.~(\ref{eq:gapincl})-(\ref{eq:gapaspect}) do not obtain, 
particularly Eq.~(\ref{eq:gapaspect}), which depends 
on the difference between $r_f$ and $r_n$, which is highly affected 
by a stellocentric offset.  

\begin{figure*}[ht]
\includegraphics[width=0.38\textwidth]{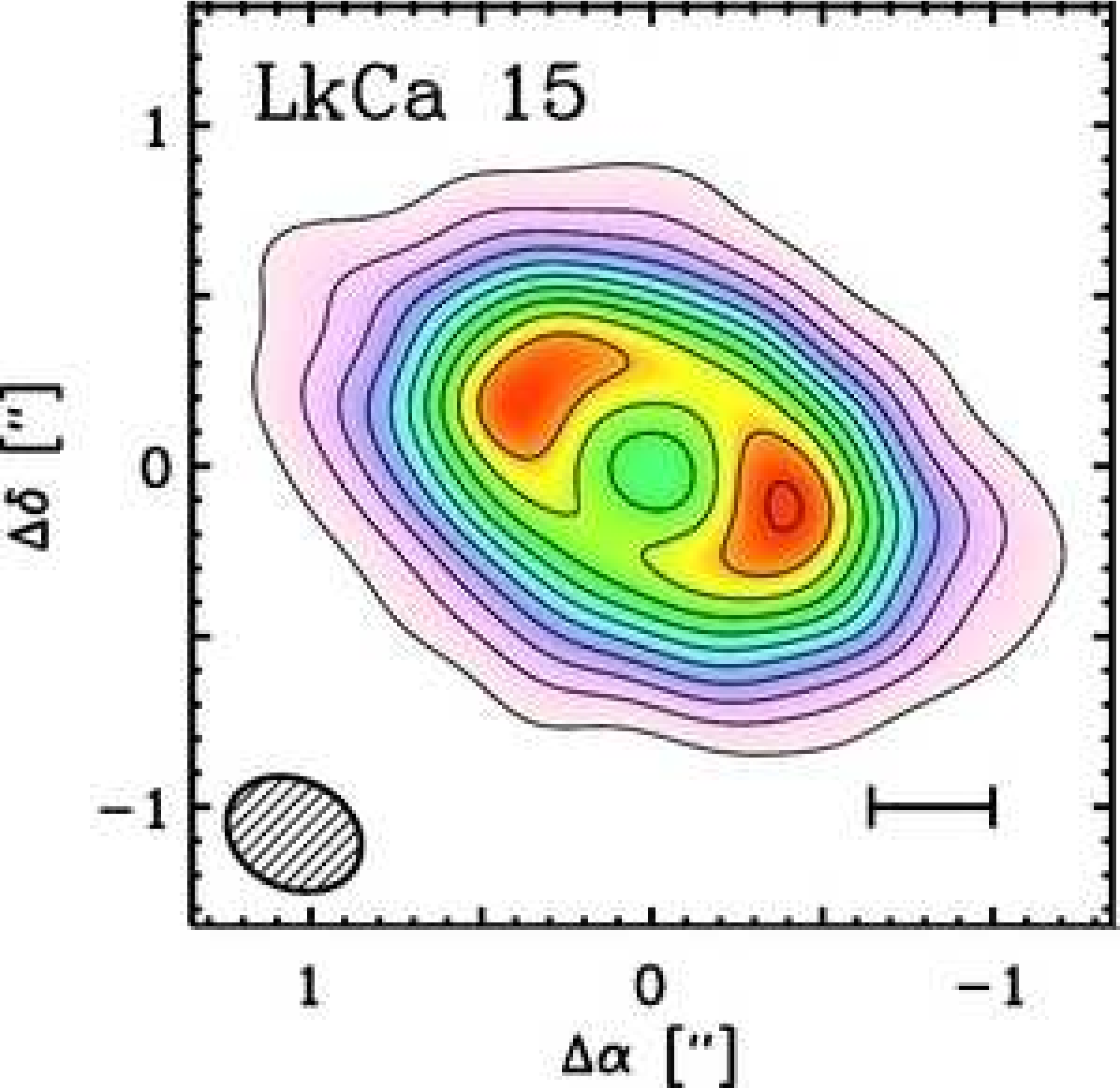}
\includegraphics[width=0.3\textwidth]{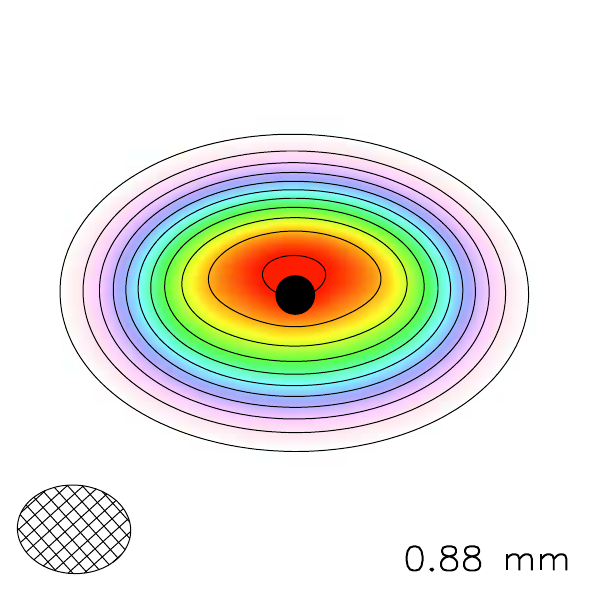}
\includegraphics[width=0.3\textwidth]{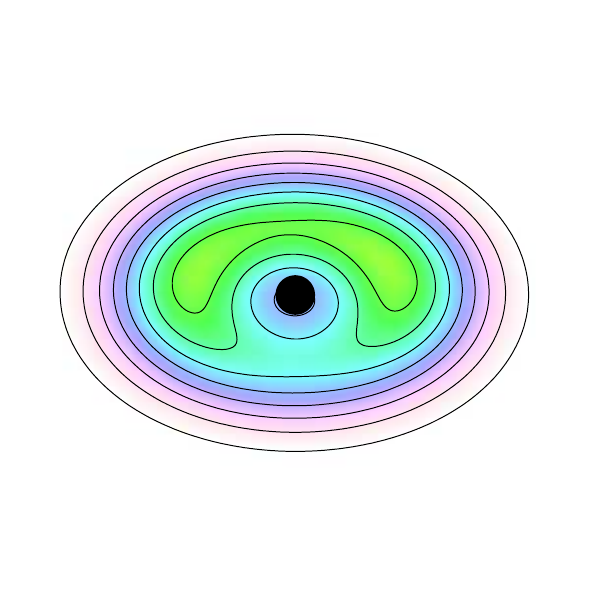}
\caption{\label{lkca15hole} 
Observed and synthetic images of the disk around LkCa15 with 
at 880 $\mu$m.  
The left panel is the SMA image obtained by \citet{2011Andrews_etal},
with the scale bar representing 50 AU, 
showing a deficit of disk material in the 
inner $\sim50$ AU of the disk.  
The center and right panels are synthetic 
images of the LkCa 15 disk modeled with a 0.5 $M_J$ planet, 
convolved with a PSF 
$0.41\arcsec\times0.32\arcsec$ in size, rotated by $-3\degr$ with respect 
to the disk major axis, matching the observations 
of LkCa 15 with the SMA \citep{2011Andrews_etal}
and indicated by the hashed ellipse.  
Contours are spaced at intervals of 
30 mJy/asec$^2$ beginning at 30 mJy/asec$^2$.  
The center panel shows the image of the disk modeled with 
a radially confined gap, with the black filled circle 
centered on the star.  
The right panel shows the image of the same disk but 
with all emission from the inner disk 
removed.  The morphology of the centrally cleared image is a 
better match to the observations, including the brightened ansae 
and far disk wall.  The right ansa is slightly brighter than the 
left, purely as a result of the orientation of the PSF.  
}
\end{figure*}

In the center panel of 
\figref{lkca15hole} we show the 0.5 $M_J$ model for LkCa 15, this time 
imaged at 880 $\mu$m.  The image has been convolved with a PSF 
matching that of \citet{2011Andrews_etal}, 
i.e.~with a beam size of $0.41\arcsec\times0.32\arcsec$ and rotated 
by $-3\degr$ with respect to the disk.  
We reproduce the flattening of the brightness profile 
in the inner region of the disk, as reported by 
\citet{2009IsellaCarpenterSargent} and \citet{2006Pietu_etal}.  
However, we do not see the inner hole resolved by 
\citet{2011Andrews_etal} and reproduced in the left panel
of \figref{lkca15hole}.  
When we simply remove the inner disk from the model and retain
only the outer disk, as shown in the right panel of \figref{lkca15hole}, 
the morphology is very close to that of \citet{2011Andrews_etal}.  
We reproduce the brightening of the far gap wall along the 
northwest minor axis, 
and the bright ansae along the major axis of the disk at the 
location of the gap wall. 
We even reproduce the brightness asymmetry along the 
southwest major axis, which results from the tilt of the 
elliptical PSF with respect to the major axis. 
This indicates that a gap in the disk is not a good model for 
LkCa 15, but rather a fully cleared inner hole, so 
if the clearing is caused by planet formation, the
planet must be more than 0.5 $M_J$ or there 
are multiple planets.

The planet candidate observed by \citet{2012KrausIreland} 
cannot have cleared all disk material out to the 
gap wall at 46 AU because it is too close to the star, 
but the observations do not rule out the presence of another planet.  
Multiple planets are capable of clearing out an inner 
hole enough so that it is optically thin, but still 
allows continued accretion onto the star
\citep{2011Dodson-RobinsonSalyk}.
A definitive measurement of the position of an additional planet in LkCa 15 
could be obtained by determining the orbital velocity of 
non-axisymmetric structure of the disk.  A planet would be 
expected to raise spiral arms in the disk, whose pattern speed would 
be that of the planet, rather than the local Keplerian orbital speed.  
Thus, a planet with semi-major axis $a$ would complete 
an orbit in $(a/1\mbox{ AU})^{3/2}$ years, 
so a feature on the inner wall of the disk at 46 AU should 
orbit at a rate of $2\arcsec\mbox{ yr}^{-1}\times(a/1\mbox{ AU})^{-3/2}$,
or 11, 9, or 8 mas/yr for a planet at 
33, 38, or 41 AU, respectively, with some variation due to 
the inclination of the disk.  

\begin{figure*}[htbp]
\plotone{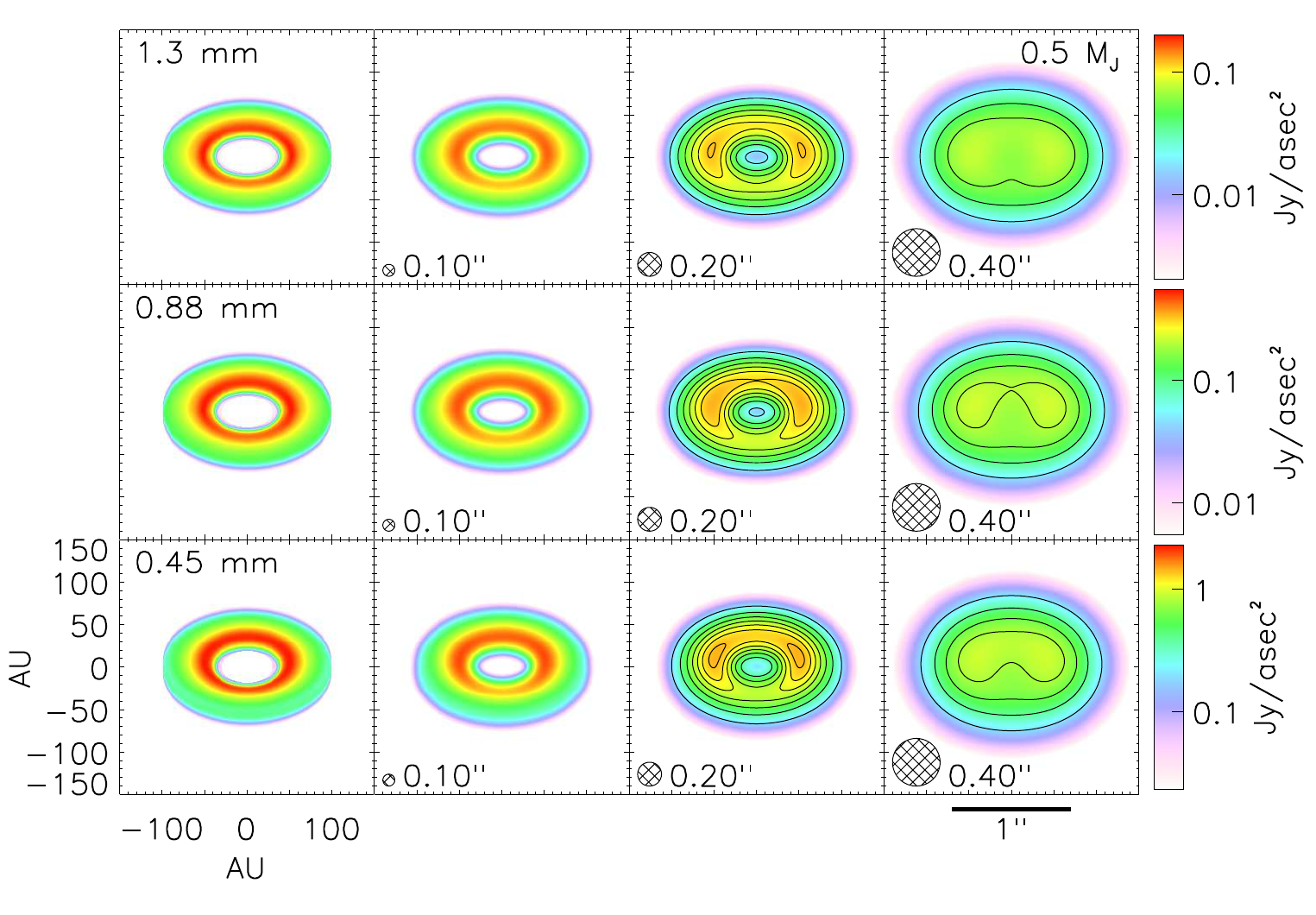}
\caption{\label{lkca15radio}
Simulated radio images of LkCa 15 with an inner hole in its disk 
at 1.3 mm (top), 0.88 mm (middle), and 0.45 mm (bottom). 
The leftmost column shows the original, fully resolved disk image.  
In the remaining columns, from left to right, the image 
has been convolved with a PSF of 0.1\arcsec, 0.2\arcsec, 
and 0.4\arcsec, respectively,
as indicated by the circles in the lower left corner of each plot.  
Contours are spaced at intervals of 
20, 50, and 200 mJy-asec$^{-2}$ at 
1.3, 0.88 and 0.45 mm, respectively.  
Contours are not shown in the left two columns for clarity.  
At these wavelengths, the far edge appears brighter at smaller angular 
resolution not because of an inherent difference in brightness, but because the 
angular size of the frontally illuminated far side of the gap is 
larger than the near side.  
}
\end{figure*}

We have shown that a gap created in a disk by a single 
planet does not successfully reproduced resolved images 
of LkCa 15, leading to the conclusion that the inner disk 
is mostly cleared out.  
In \figref{lkca15radio}, we show predicted radio images of 
the LkCa 15 disk model with a cleared inner hole, from 
the right panel for \figref{lkca15hole}.  
The images are continuum emission at 1.3, 0.88, and 0.45 mm, 
wavelengths that ALMA is able to probe.  
The images are convolved with PSFs of varying sizes to illustrate 
how the apparent morphology of the disk changes with 
different beam sizes.  If the beam size is larger than the angular 
size of the brightened inner wall, the brightness of the feature 
will be diminished and diluted.  These images demonstrate that 
ALMA should be able to clearly resolve the inner hole 
in LkCa 15, and determine just how empty the inner clearing is.  

The inner hole model shown above is a lower limit on the brightness 
of the inner wall.  The model is based on a disk model with a 
deep, wide gap, and then simply subtracting emission from the inner disk.  
The model was calculated assuming the presence of an inner disk, 
which would prevent full illumination of the inner wall of a cleared 
hole.  In a disk with a true inner hole, the wall will be hotter and 
therefore brighter.  The brightened wall exposed by the inner clearing 
may partially shadow the disk material behind it, 
creating still more contrast 
between the inner wall and the disk outside it.

\section{Discussion and Conclusions}

The gaps modeled in this paper are based on a simple parameterized 
structure for the gap.  Real gaps created by planets in 
disks may show non-axisymmetric structure not captured in this 
simple model.  Thus, while the models and simulated images 
presented in this work give a good qualitative description 
of the effects of gap-opening on disks, the magnitude of 
shadowing and illumination effects may vary depending on the exact 
disk properties.  

Since the models presented here do not include hydrodynamics, 
the effects of accretion onto the embedded planet have been neglected.  
We can estimate the accretion rates from the empirical 
determination of \citet{bate}, 
\begin{equation}
\dot{M_p} = b \frac{M_p}{M_*} \rho \Omega_p a^3
\end{equation}
where $b\approx2.3$.  
For the 70 and 200 $M_{\oplus}$ planets modeled here, the estimated
accretion rate is about $\sim 10^{-4}\,M_J\mbox{ yr}^{-1}$.  
We can estimate the accretion luminosity as 
\begin{equation}
L\sub{acc} = \frac{G_{gr} M_p \dot{M_p}}{r\sub{Hill}}
\end{equation}
where $G_{gr}$ is the gravitational constant.  
The Hill radius is 0.4 and 0.6 AU for the 70 and 200 
$M_{\oplus}$ planet, respectively.  
Then the accretion luminosity is $3\times10^{28}$ 
and $6\times10^{28}$ erg s$^{-1}$, respectively.  
Assuming that the material interior to the Hill sphere is 
optically thick, then the effective temperature may be 
expressed as 
$ T\sub{eff} = (L\sub{acc}/(4\pi r\sub{Hill}^2 \sigma_B) )^{1/4}. $
The resultant temperature is about $\sim30$ K for both 
planet masses, while the midplane temperature in the absence 
of this accretion heating is 46 K and 39 K for the 
70 and 200 $M_{\oplus}$ planet, respectively.  
The additional heating may be expressed as 
$T=(T_1^4+T_2^4)^{1/4}$, so the addition of the accretion 
luminosity would raise the temperature immediately around the planet 
by a few degrees.  
The modest amount of heating suggests that little to no detectable 
change in the structure of the disk would result.  
The amount of heating could be increased 
if the radius of accretion were less than the Hill radius, but 
that calculation is outside the scope of this paper.  
In any case, the heating from the planet would be a local 
asymmetric perturbation on the disk structure and the 
structure of the gap upstream and downstream of 
the planet would be unchanged.  

We have shown that the geometric effects of inclined disks, with 
or without gaps, can create structure in disk images that must 
be interpreted carefully.  The effect of inclination on disk images 
varies with wavelength, since scattered light images in the optical 
or infrared probe only the disk surface, while radio frequency images 
probe the deep thermal structure of the disk.  

The scattered light models here (1 $\mu$m images) apply over a 
limited range of inclination angles, in that the brightness calculated 
is that scattered off the surface of the disk, rather than 
light seen extinguished through the disk.  
That is, our models do not apply to a system such as HH 30, 
which is nearly edge on \citep{2001Cotera_etal}, so that 
the observed scattered light is seen through the disk.  

Comparing brightness profiles along different 
axes can yield the inclination and aspect ratio of the disk, 
under the assumption of axisymmetry.  This can be done, in principle,
at any wavelength, whether in scattered light or thermal emission, 
and with or without a gap in the disk.  

When a gap is opened in an inclined disk, the near side of the gap's
outer wall appears slightly brighter than the far side at 
the point of maximum brightness.  However, if the 
illuminated wall is unresolved in the observation, 
then the far side may appear brighter because it 
spans a larger angular size.  
A gap in a disk may aid in the calculation of inclination angle 
and aspect ratio of a disk, but only if the gap is not eccentric.  
An eccentric gap, possibly created by an embedded planet, affects the 
derivation of aspect ratio more than the inclination angle, 
as shown in the example of LkCa 15.  

Our example of LkCa 15 demonstrates that gaps in disks are currently 
detectable in scattered light.  However, the gap in LkCa15 is 
at a much larger radius ($\sim50$ AU) compared to the gap at 10 AU 
as modeled in the first part of this paper.  
Supposing that the 10 AU gap was in a disk at 
140 pc, the distance of Taurus, would the gap be detectable?  
The gaps modeled were 1.1 and 1.7 AU in width, respectively, or 
8 and 12 mas, respectively.  Observing at 1 micron at the 
diffraction limit, this would require a $17-27$ m telescope to resolve.  
At 0.3 mm, the baseline required would be $5-8$ km.  
Large optical telescopes such as the LBT, GMT, and TMT, and 
the radio array ALMA would achieve such resolving power, 
but prospects for imaging gaps at mid- to far-infrared wavelengths 
are small.  In the optical, high contrast imaging would also be necessary.  
An inner working angle of 0\arcsec.05 would block out the inner 
7 AU of the disk.  At sub-mm to mm wavelengths, stellar contrast 
is not an issue, but the contrast within the gap itself is much less 
than at shorter wavelengths.  
At 0.3 and 1 mm, the required sensitivity is on order of 1 and 0.1 
Jy/asec$^2$, or 1.4 and 0.15 K
respectively.  In band 9 of ALMA (0.45 mm), the required integration 
time for 10 mas resolution with 1.4 K sensitivity using 50 antennas
is 22 hours.  At 1 mm (band 7), 0.15 K sensitivity 
requires 13 hours.  

By comparing resolved images of disks 
to the models presented here, we can estimate the masses of 
planets that might be causing those gaps.  
In particular, we put a lower mass limit of 0.5 Jupiter masses 
on a planetary companion in LkCa 15 that would create the 
observed inner hole.  
If planets are responsible for the inner hole, then 
our results suggest that it is either caused by a more massive planet 
or by multiple planets with overlapping gaps.

\acknowledgements
The authors thank C.~A.~Grady and A.~Hubbard 
for helpful discussions in the preparation of this paper.  
We also thank an anonymous referee for constructive comments that 
greatly improved this paper.    
H.J.-C. acknowledges support from the NASA Astrophysics Theory Program 
through grant NNX12AD43G and the 
the Michelson Fellowship Program under contract with the
Jet Propulsion Laboratory (JPL) funded by NASA.  N.J.T. 
was employed by JPL, which is managed for NASA by the California
Institute of Technology.  He was supported by the NASA Origins of
Solar Systems program through grant 09-SSO09-0046, and by the
Humboldt Foundation through a Fellowship for Experienced
Researchers.

\appendix

\section{Scattering Angles}
\label{app:angles}

In this Appendix, we show how to calculate $\eta$, 
the angle between the surface normal and the observer, assuming 
an axisymmetric disk. 
For a general surface, $z = f(x,y)$, the unit surface normal can
be expressed as 
\[
\hat{\mathbf{n}} = \frac{-(\partial z/\partial x) \hat{\mathbf{x}} 
- (\partial z/\partial y) \hat{\mathbf{y}} + \hat{\mathbf{z}}}{
\sqrt{(\partial z/\partial x)^2 + (\partial z/\partial y)^2 + 1}}
\]
Without loss of generality, we place the disk midplane in the $xy$ plane 
and the observer in the 1st quadrant of the $xz$ plane.  
Then, if the inclination angle 
is $i$, the vector toward the observer is 
$\hat{\mathbf{m}}= \sin i \,\hat{\mathbf{x}} + \cos i \,\hat{\mathbf{z}}$.
Then the cosine of the angle between the surface normal and observer is 
\[ \cos\eta = \hat{\mathbf{n}} \cdot \hat{\mathbf{m}}
= 
\frac{-(\partial z/\partial x) \sin i
+\cos i
}{
\sqrt{(\partial z/\partial x)^2 + (\partial z/\partial y)^2 + 1}}
\]
We convert this to cylindrical coordinates, with 
\begin{eqnarray}
x &=& r \cos\theta \nonumber\\
y &=& r \sin\theta \nonumber\\
z &=& z \nonumber
\end{eqnarray}
and using the chain rule to get
\begin{eqnarray}
\frac{\partial z}{\partial x} 
&=& \cos\theta\frac{\partial z}{\partial r} 
- \frac{\sin\theta}{r}\frac{\partial z}{\partial \theta} \nonumber\\
\frac{\partial z}{\partial y}
&=& \sin\theta\frac{\partial z}{\partial r}
+ \frac{\cos\theta}{r}\frac{\partial z}{\partial \theta} \nonumber\\
\end{eqnarray}
and find
\begin{equation}
\cos\eta = 
\frac{\cos i-\left[\cos\theta (\partial z/\partial r)
+ (\sin\theta/r)(\partial z/\partial \theta)\right]\sin i}{
\sqrt{(\partial z/\partial r)^2 + (1/r^2)(\partial z/\partial\theta)^2 + 1}}
.\end{equation}
In an axisymmetric disk, $\partial z/\partial\theta=0$.  
Defining $\tan\alpha = \partial z/\partial r$, 
\begin{equation}
\cos\eta =\cos\alpha\cos i-\cos\theta\sin\alpha\sin i.
\end{equation}
Since the observer is located toward positive $x$, 
the far side of the disk is $\theta=\pi$ and 
the near side of the disk is $\theta=0$.  
Hence, on the far side of the disk, 
$\eta_1=i-\alpha$ and on the near side $\eta_2=i+\alpha$.  
At maximum elongation, $\theta=\pm\pi/2$ and 
$\cos\eta=\cos\alpha\cos i$.  

\bibliographystyle{apj}
\bibliography{../../../planets,../../../jang-condell,LkCa15,gaps}

\end{document}